\documentclass[usenatbib,useAMS,usegraphicx,fleqn]{mn2e}
\usepackage{url}
\usepackage{amsmath}
\usepackage{wasysym}
\usepackage{txfonts}
\usepackage{mathrsfs}
\usepackage{bm}

\newcommand\apj{ApJ}
\newcommand\apjl{ApJ}
\newcommand\aj{AJ}
\newcommand\mnras{MNRAS}
\newcommand\aap{A\&A}
\newcommand\nat{Nature}

\renewcommand\pi\upi
\newcommand\abs[1]{\left\lvert#1\right\rvert}
\newcommand\z{\bar z}
\newcommand\e{\mathrm e}
\newcommand\df{\mathrm d}
\newcommand\ij{\mathrm i}
\renewcommand\partial\upartial

\newcommand\squ\square
\newcommand\tri\triangle
\newcommand\hex\hexagon
\newcommand\ssqu{\scriptsize\squ}
\newcommand\stri{\scriptsize\tri}
\newcommand\shex{\scriptsize\hex}

\DeclareMathOperator{\sgn}{sgn}

\DeclareMathOperator{\sn}{sn}
\DeclareMathOperator{\cn}{cn}
\DeclareMathOperator{\dn}{dn}

\pagerange{\pageref{start}--\pageref{finish}}
\pubyear{2007}

\journal{{\sc J.~An}:
{\it Gravitational lensing by point masses on regular grid points}}
\date{
{\sc to appear in}
{\sl Monthly Notices of the Royal Astronomical Society}}

\title[Lensing by regular point-mass lattices]
{Gravitational lensing by point masses on regular grid points}

\author[J.~An]{Jin H. An\thanks{E-mail: jinan@space.mit.edu}\\
MIT Kavli Institute for Astrophysics \& Space Research,
Massachusetts Institute of Technology, 77 Massachusetts Avenue,
Cambridge MA 02139, USA}

\begin{document}
\label{start}
\maketitle

\begin{abstract}
It is shown that gravitational lensing by point masses arranged in
an infinitely extended regular lattice can be studied
analytically using the Weierstrass functions.
In particular, we draw the critical curves and
the caustic networks for the lenses arranged in regular-polygonal --
square, equilateral triangle, regular hexagon -- grids. From this,
the mean number of positive parity images as a function of the average
optical depth is derived and compared to the case of the infinitely
extended field of randomly distributed lenses. We find that the high
degree of the symmetry in the lattice arrangement leads to a
significant bias towards canceling of the shear caused by the
neighboring lenses on a given lens position and lensing behaviour
that is qualitatively distinct from the random star field. We also
discuss some possible connections to more realistic lensing scenarios.
\end{abstract}
\nokeywords

\section{Introduction}

Beyond the Galactic and Local-Group (in particular, Magellanic Clouds
and M31) microlensing experiments \citep{Al00,Af03,CN05,dJ06}, the
main interest on the point-mass lenses lies in the effect of the
individual stars in a lensing galaxy
\citep{CR79,CR84,NO84,Ir89,WMS95,SW02,Ke06}.
For these cases, each point-mass lens cannot be treated
individually and their collective effects usually differ drastically
from that of the simple linear superposition of them.
The usual approach to this ``high-optical-depth microlensing'' problem is
``inverse ray tracing'' \citep{KRS,SW87,WPS,WWS}, that is, examining
statistical properties of lensing observables using Monte-Carlo
realizations of the lensing system. However, as
the number of the individual point-mass lenses to reproduce
the realistic scenario well approaches the number of stars in the galaxy,
this becomes very expensive rather quickly in terms of the required
resources. Another complementary
approach to the problem is from the random walk and the probability
theory together with the thermodynamic approximation
(that is, effectively considering the limit at the infinite number
of the stars), which has been quite successful in certain
well-defined problems \citep{NO84,Sc87,SEF}.

In the following, we consider a different approach to the problem:
seeking complete analytic solutions. In many complex
systems, imposing high level of symmetry can reduce the problem to a
simpler one, which may be otherwise too complicated to analyze. In
a similar spirit, in this paper we consider gravitational
lensing by equal point masses that are arranged in an infinitely
extended regular lattice. After we derive the appropriate lens
equation in Sect.~\ref{sec:le}, they are examined in Sect.~\ref{sec:cc} to
find the corresponding critical curves and the caustic networks. In
Sect.~\ref{sec:mn}, we study the mean number of positive parity
images, which is compared to the case of lensing by an infinite
random star field. In Sect.~\ref{sec:sh}, we consider the effect of
the external potential by mean of adding external shear
(and convergence), and discuss some possible
connections to more realistic lensing scenarios.

We argue that, while the system considered here may be somewhat artificial
in its construct and rather abstract in its nature, the study presented
in this paper can lead us to some insight on microlensing at high optical
depth.

\section{Lens equation}
\label{sec:le}

In most astronomical situations, gravitational lensing is
described by the lens equation
%
$\bm y=\bm x-\bm\nabla\psi$,
%
which is derived from the lowest-order nontrivial approximation to the
path of light propagation \citep{SEF,PLW,KSW}. Here, $\bm y$ and $\bm x$
are the (2-dimensional) vectors representing the lines of sight toward,
respectively, the angular positions of the source under the absence of
lensing and the lensed image. The lensing potential $\psi$ is the
gravitational potential (equivalently, the gravitational `Shapiro'
time delay) integrated along the line of sight up to a scale
constant \citep{BN86}.

In particular, gravitational lensing by a point mass is described
by the potential $\psi(\bm x)=\theta_\mathrm E^2 \ln\abs{\bm x-\bm l}$
if the point-mass lens is located along the line of sight indicated by
the vector $\bm l$ \citep{Pa86}. Here, the scale constant $\theta_\mathrm E$
is known as the `Einstein (ring) radius' and determined by the mass of
the lens ($\theta_\mathrm E^2\propto M$) and the distances among the
source, the lens, and the observer.
With this potential, we find the lens equation for point-mass lensing,
\begin{equation}
\bm y=\bm x-\frac{\bm x-\bm l}{\abs{\bm x-\bm l}^2}
\end{equation}
where the unit of angular measurements has been chosen such that
$\theta_\mathrm E=1$. The equation is generalized to the case of
multiple point-mass lenses with a smoothly-varying `large scale'
potential \citep{Yo81} such that
\begin{equation}
\label{eq:leq}
\bm y=\bm x-\sum_{i=1}^N\frac{m_i}M
\frac{\bm x-\bm l_i}{\abs{\bm x-\bm l_i}^2}
-\bm\nabla\psi_\mathrm{ext}.
\end{equation}
Here, the unit of angular measurements is now given by $\theta_\mathrm E$
corresponding to a fiducial mass scale $M$ and subsequently the
`external' lensing potential $\psi_\mathrm{ext}$ should be rescaled
appropriately. In addition, the masses of individual lenses $m_i$ enter
into the equation as the ratio to the fiducial scale.


Next, let us think of the case that the equal point-mass lenses are
located at the lattice points on an infinitely extended square
grid. In addition, we ignore the effect of the external potential
for the moment. Utilizing the complex-number notation
\citep{BK75,SCN,Wi90,An05}, the lens equation (\ref{eq:leq}) then can
be written as
\begin{equation}
\label{eq:cdef}
\begin{split}&
\eta=z-\overline{f(z)}
\,;\\&
f(z)=\sum_i\frac1{z-\lambda_i}
=\sum_{m,n=-\infty}^\infty\frac1{z-2(m+n\ij)\omega},
\end{split}
\end{equation}
where $\eta$, $z$, and $\lambda$ are the complexified variables
corresponding to $\bm y$, $\bm x$, and $\bm l$, respectively, and the
over-bar notation is used to indicate complex conjugation. We further
assume that the lens locations are given by $\lambda=2(m+n\ij)\omega$
where $m$ and $n$ are any pair of integers (both running from the negative
infinity to the positive infinity) and $\omega$ is the half distance
between adjacent grid points. Without any loss of generality, $\omega$
can be restricted to be a positive real. Strictly speaking, the
`deflection function' $f(z)$ given formally in terms of the infinite
sum in equation (\ref{eq:cdef}) is not well-defined as it is not convergent.
This is in fact a natural consequence of having an infinite total lensing
mass, which is unphysical. However, what is important in our
understanding of the lensing is not the absolute amount of the deflection
but the difference of the deflections between neighbouring lines of sights.
That is to say, if we add any constant to the deflection function
in the lens equation, the resulting lens equation is reduced to the
equivalent one without the constant by introducing the ``offset'' to
the source position to cancel the constant. Two equations then are
observationally equivalent because there is no a priori information
regarding the source position in the absence of the lens.
Similarly, despite the seeming global failure of equation (\ref{eq:cdef}),
we can still proceed by focusing on the local properties of the lensing
described by equation (\ref{eq:cdef}).

For this, we first write down the (formal) second-order complex derivative
of the deflection function;
\begin{align}
\begin{split}
f''(z)&
=\sum_{m,n=-\infty}^\infty\frac2{[z-2(m+n\ij)\omega]^3}
\\
\equiv&-\wp'(z\vert\omega,\ij\omega)=-\wp'\left(z;g_2,0\right)
\end{split}\\
\intertext{where}
\label{eq:g2}
g_2
=&\left(\frac{\omega_0}\omega\right)^4
=\frac{\Gamma^8_{1/4}}{16\pi^4}\kappa_\star^2.
\end{align}
We find that this is related to the definition of the special function
known as the Weierstrass elliptic function\footnote{\scriptsize Throughout this
paper, the arguments of the Weierstrass functions followed by a
vertical bar -- $\vert$ -- denote the half-periods whereas the
elliptic invariants are indicated by those followed by a semicolon
(;). For details, see the listed references or Appendix. Whenever it
is appropriate, these arguments may be suppressed, provided that there
is little danger of confusion.} \citep[or see 
also Appendix]{AS}. Here, $\Gamma_x\equiv\Gamma(x)$ is the gamma function,
$\wp(z)$ is the Weierstrass elliptic
function and $\wp'(z)$ is its complex derivative. We also use
an abbreviated notation for (the power to) the gamma function such that
$[\Gamma(x)]^n=\Gamma^n_x$ in order to avoid proliferation of
parentheses and brackets. In addition,
$\omega_0\approx1.854$ is the real half-period of the lemniscatic case
of $\wp$-function (see Appendix) and $\kappa_\star=\pi/(2\omega)^2$ is the
optical depth \citep{VO83,Pa86} of the lenses. If we consider the
square cell given by the centres of each lensing lattice, it is
straightforward to see that there exists a lens corresponding to each
cell with side length of $2\omega$. Hence, the mean surface density of
point masses is in fact given by $\sigma=(2\omega)^{-2}$, and therefore,
the optical depth by $\kappa_\star=\pi\sigma=\pi/(2\omega)^2$.
Moreover, the 90\degr-rotational symmetry of the system implies that
the total shear due to all the other lenses on any given lens cancels
out to be zero [mathematically this indicates the constant term in the
Laurent-series expansion of $f'(z)$ is null]. Consequently, we find
that $f'(z)=-\wp(z;g_2,0)$ and $f(z)=\zeta(z;g_2,0)+C$. Here,
$\zeta(z)$ is the Weierstrass zeta function \citep{AS}
and $C$ is an arbitrary complex constant. If we choose the `centre' of
the system to coincide with the lens at the coordinate origin, then we
have $C=0$. We note that, while the constant $C$ cannot be
independently determined, its choice is physically inconsequential.
The situation is analogous to lensing by an infinitely extended
mass screen. While a na\"ive expectation from the symmetry appears to
indicate null deflections along every line of sight, detailed
physical consideration leads to the uniform focusing with respect to
an arbitrary choice of the centre. In the current situation, the
discrete translation symmetry implies the arbitrariness in the choice
of the centre and consequently that of $C$, which results in the
infinite sum in equation (\ref{eq:cdef}) not being well-determined.
However, the constant-deflection term in the lens equation again only leads
to a constant offset between the source and the image/lens plane,
which has no observable consequence and therefore can be ignored.

Some results are immediate from the resulting lens equation
\begin{equation}
\label{eq:le}
\eta=z-\overline{\zeta(z;g_2,0)}.
\end{equation}
For instance, the property of $\zeta(z)$ such that
\begin{equation}
\zeta\bigl[z+2(m+n\ij)\omega;g_2,0\bigr]=
\zeta(z;g_2,0)+\frac{\pi}{2\omega}\,(m-n\ij)
\end{equation}
for arbitrary integers $m$ and $n$ indicates that the vertices $z$,
$z+2\omega$, $z+2(1+\ij)\omega$, and $z+2\ij\omega$ of a square in the
lens plane map to the points on the source plane:
$\eta=z-\overline{\zeta(z)}$, $\eta+\aleph$, $\eta+(1+\ij)\aleph$, and
$\eta+\ij\aleph$ where $\aleph=2\omega-\pi/(2\omega)$, and therefore the
mean inverse magnification is given by
$[\aleph/(2\omega)]^2=[1-\pi/(2\omega)^2]^2$. The result confirms the
expectation that the sufficiently large beam of light would see the
system as if it were a uniform screen of mass with a convergence (or
the optical depth) $\kappa=\pi\sigma=\pi/(2\omega)^2$. Along the
real line, equation (\ref{eq:le}) maps each segment between two
adjacent lenses onto the whole real line, and thus it is easily argued
that there are infinite number of images. However, unless
$\kappa_\star=1$, all but a finite number of images have negative parity.

\input f1.cap

\subsection{lenses on triangular and hexagonal grid}

Analogous to the preceding case, we can also set up the lens equation
for the equal point-mass lenses located at the lattice points on an
infinitely extended equilateral-triangular grid. The
60\degr-rotational symmetry of the system again implies that the total
shear on any given lens caused by the remaining set of lenses also
cancels out. This fact, combined with the general definition of
$\wp$-function indicates that the lens equation can again be written
down using the Weierstrass functions. If the lenses are placed over
the grid given by $\lambda=2\omega(m+n\e^{\pi\ij/3})$ where $m$ and $n$
are integers, the corresponding lens equation is given by
\begin{gather}
\label{eq:leq_tri}
\eta=z-\overline{f(z)}\,;\qquad
f(z)=\zeta(z\vert\omega,\e^{\pi\ij/3}\omega)=\zeta(z;0,g_3)
\\
\intertext{where}
\label{eq:g3}
g_3=\left(\frac{\omega_2}\omega\right)^6
=\left(\frac{\sqrt3\Gamma^6_{1/3}}{8\pi^3}\right)^3\kappa_\star^3
\end{gather}
and $\omega_2\approx1.530$ is the real half-periods of the
equiharmonic case of $\wp$-function (see Appendix) whereas
$\kappa_\star=\pi/(2\sqrt3\omega^2)$ is the corresponding optical
depth. Likewise, the quasiperiodicity of $\zeta(z;0,g_3)$,
\begin{equation}
\zeta\bigl[z+2(m+n\e^{\pi\ij/3})\omega;0,g_3\bigr]=
\zeta(z;0,g_3)+\frac\pi{\sqrt3\omega}\,\bigl(m+n\e^{-\pi\ij/3}\bigr)
\end{equation}
where $m$ and $n$ are arbitrary integers, indicates that the vertex
points $z$, $z+2\omega$, and $z+2\e^{\pi\ij/3}\omega$ of the equilateral
triangle in the lens plane map onto the source plane points:
$\eta=z-\overline{\zeta(z)}$, $\eta+\beth$ and $\eta+\beth\e^{\pi\ij/3}$
where $\beth=2\omega-\pi/(\sqrt3\omega)$ so that the mean inverse
magnification is again given by $(1-\kappa_\star)^2$ where the optical
depth or the mean convergence is $\kappa_\star=\pi/(2\sqrt3\omega^2)$
(and the mean surface density $\sigma$ of the point masses on the
triangular grid is given by $\sigma^{-1}=2\sqrt3\omega^2$ where
$2\omega$ is the side length of the unit triangular cell).

We note that the the preceding two cases of the lens arrangement
correspond to two of the three possible regular tessellations of the
two-dimensional plane. It may be of some interest to consider the
regular lens placement corresponding to the remaining regular
tessellation -- the hexagonal or honeycomb tiling. It turns out that
the regular hexagonal grid case is closely related to the equilateral
triangular grid.\footnote{\scriptsize Note that they are in fact so-called dual tiling
of each other. That is, if we consider a grid with vertex points at the
centres of each triangular cell of the equilateral triangular grid,
then the resulting lattice forms a regular-hexagonal (honeycomb) grid
and also vice versa.} By removing every third of the lens that forms a
(30\degr-rotated) larger triangular grid of side length of
$2\sqrt3\omega$ from the base triangular grid with side length of
$2\omega$, the remaining lenses form a hexagonal grid (or the vertices
of honeycomb cells). Consequently, the corresponding lens equation may
be written as
\begin{equation}
\label{eq:leq_hex}
\begin{split}&
\eta=z-\overline{\mathfrak d(z;g_3)}
\,;\\&
\mathfrak d(z;g_3)=
\zeta(z\vert\omega,\e^{\pi\ij/3}\omega)
-\zeta(z\vert\sqrt3\e^{\pi\ij/6}\omega,\sqrt3\ij\omega)
\\&\qquad
=\zeta(z;0,g_3)-\zeta\Bigl(z;0,-\frac{g_3}{27}\Bigr)
\end{split}
\end{equation}
where $g_3$ is related to $\omega$ through equation (\ref{eq:g3})
provided that $2\omega$ is still the side length of the base
triangular grid (and also that of the unit honeycomb cell). However,
since every third of lens has been removed from the base triangular
cell, the optical depth is reduced to
$\kappa_\star=\pi/(3\sqrt3\omega^2)$, and thus
\begin{equation}
g_3=\left(\frac{3^{3/2}\Gamma^6_{1/3}}{2^4\pi^3}\right)^3\kappa_\star^3.
\end{equation}
In addition, $\mathfrak d'(z;g_3)=-\mathfrak h(z;g_3)$ (the primed
symbol indicates the complex differentiation, i.e., the derivative
with respect to $z$). The function $\mathfrak h(z;g_3)$ is studied in
Appendix \ref{app:hex}. It is again straightforward to establish that
the mean inverse magnification is $(1-\kappa_\star)^2$. Unlike the
preceding two cases, the system described by lens equation
(\ref{eq:leq_hex}) does not have the lens at the centre, but the
centre of the system corresponds to the location on the lens plane
such that $\mathfrak d'(0)=\mathfrak h(0)=0$. However, we note that
this is deliberately chosen for mathematical simplicity, and does
not imply any intrinsic difference of the honeycomb grid from the
square or triangular grid. In particular, we note the
120\degr-rotational symmetry of the system with respect to any lens
location, which actually implies that any lens on the honeycomb grid
experiences zero shear from the remaining point masses. In other
words, for all three cases of the point masses on the regular grid,
the total shear from all the adjacent lenses exactly cancels out at
any lens position.

\input f2.cap

\input f3.cap

\subsection{lensing potential}

\input f4.cap

While most of the lensing properties of the lattice lens can be
studied using the lens equations, it is still useful to have an
expression for the lensing potential for some purposes such as the time
delay. Although the direct infinite sum of the individual logarithmic
potential of the point-mass lens is divergent everywhere, one can
get around this difficulty through the known antiderivative of
the Weierstrass zeta function. First, we note the relation between the
real potential and the complex deflection function \citep{An05,AE06};
the complex lens equation is given by $\eta=z-2\partial_{\z}\psi$
if the real potential is given by $\psi$. Here, the operator
$\partial_{\z}$ indicates the `Wirtinger derivative'
\citep[e.g.,][]{SK95} with respect to $\z$, that is, if
$\psi=\psi(x,y)$ and $z=x+y\ij$, then
\begin{equation}
\partial_{\z}\psi
=\frac{\partial\psi}{\partial x}\frac{\partial x}{\partial\z}
+\frac{\partial\psi}{\partial y}\frac{\partial y}{\partial\z}
=\frac12\frac{\partial\psi}{\partial x}
+\frac\ij2\frac{\partial\psi}{\partial y}
\end{equation}
because $x=(z+\z)/2$ and $y=\ij(\z-z)/2$. For the current
scenario, we have $f(z)=2\overline{\partial_{\z}\psi}=
2\partial_z\bar\psi=2\partial_z\psi$ (note that $\psi=\bar\psi$ since
$\psi$ is real). Moreover, $f(z)$ is a complex analytic function and
so there exists a complex analytic function $\psi_c(z)$ such that
$f(z)=\psi_c'(z)=\partial_z\psi_c$ and $\partial_{\z}\psi_c=0$.
Then, $\partial_z(\psi_c+\bar\psi_c)=\partial_z\psi_c+\partial_z\bar\psi_c
=\partial_z\psi_c+\overline{\partial_{\z}\psi_c}=f(z)$, and thus,
$2\psi=\psi_c+\bar\psi_c$, that is, $\psi(x,y)=\Re[\psi_c(x+y\ij)]$ up
to an additive constant \citep{An05}. Here, $\Re[f]$ is the real
part of a complex-valued function $f$.
Consequently, we find that
the real potential up to an additive constant for the regular lensing
lattice is given by
\begin{equation}
\label{eq:pot}
\psi(x,y)=
\begin{cases}
\ln\abs{\sigma(z;g_2,0)}&\squ\\
\ln\abs{\sigma(z;0,g_3)}&\tri\\
\ln\abs{\dfrac{\sigma(z;0,g_3)}{\sigma(z;0,-g_3/27)}}&\hex
\end{cases}
\end{equation}
where $z=x+y\ij$ and $\sigma(z;g_2,g_3)$ is the Weierstrass sigma
function \citep{AS},
which can be defined as the anti-log-derivative of the Weierstrass
zeta function, that is, $(\df/\df z)\ln\sigma(z)=\zeta(z)$. Here, we have
also used the property of the complex logarithm function such that
$\Re[\ln g]=\ln\abs{g}$ where $g$ is any complex-valued function. Note
that an additional linear term may be required to be added in the
expression for the potential if the centre of the system is chosen
differently from those of the lens equations discussed earlier.
Fig.~\ref{fig:pot} shows contour plots for the equipotential lines for the
potential given in equations (\ref{eq:pot}).

\section{Critical curves and Caustics}
\label{sec:cc}

As usual, the Jacobian determinant of equation (\ref{eq:le}),
(\ref{eq:leq_tri}) or (\ref{eq:leq_hex})
\begin{equation}
\mathscr J=1-\abs{f'(z)}^2=
\begin{cases}
1-\abs{\wp(z;g_2,0)}^2&\squ\\
1-\abs{\wp(z;0;g_3)}^2&\tri\\
1-\abs{\mathfrak h(z;g_3)}^2&\hex
\end{cases}
\end{equation}
is the inverse magnification of the individual image at the limit of
the point source. In addition, the critical curves can be found from
$\abs{f'(z)}=1$ or equivalently solving $f'(z)=-\e^{-2\ij\phi}$ for
$z(\phi)$ with $\phi$ being a real parameter
\citep{Wi90,WWS,AE06},
and the caustics from their images under the
mapping given by the lens equation. Since $f'(z)$ is elliptic (hence,
biperiodic), there exists an infinite set of critical curves and
caustics although they can all be recovered from the discrete grid
translation of a `unit' curve.

The homogeneity relation of $\wp$-function indicates that
\begin{equation}
\begin{split}
\wp(z;g_2,0)&
=\left(\frac{\omega_0}\omega\right)^2
\wp\Bigl(\frac{\omega_0}\omega\,z;1,0\Bigr)
\,;\\
\wp(z;0,g_3)&
=\left(\frac{\omega_2}\omega\right)^2
\wp\Bigl(\frac{\omega_2}\omega\,z;0,1\Bigr)
\end{split}
\end{equation}
and from equation (\ref{eq:hfun}) that
\begin{equation}
\mathfrak h(z;g_3)
=\frac{\e^{-2\pi\ij/3}}3\left(\frac{\omega_2}\omega\right)^2
\Biggl[\wp\biggl(\frac{\e^{\pi\ij/6}}{\sqrt3}\frac{\omega_2}{\omega}z;
0,1\biggr)\Biggr]^{-2}.
\end{equation}
Hence, the topography of the critical curves may be studied from the
lines of constant values of $\abs{\wp(z;1,0)}$ (square grid) and
$\abs{\wp(z;0,1)}$ (triangular and hexagonal grid). In particular, the
critical curves for the square (or triangular) lens lattice with the
half-period $\omega$ are basically the curves of constant
$\abs{\wp(z;1,0)}=\omega^2/\omega_0^2$ [or
$\abs{\wp(z;0,1)}=\omega^2/\omega_2^2$] rescaled by a factor of
$\omega/\omega_0$ (or $\omega/\omega_2$). On the other hand, those for
the honeycomb lattice are found from the curves of constant
$\abs{\wp(z;0,1)}=\omega_2/(\sqrt3\omega)$ scaled by
$\sqrt3\omega/\omega_2$ and rotated by $-30\degr$. Since
$\wp$-functions are well-defined elliptic functions, for all three
cases the critical curves are an infinite set of nonintersecting
closed curves except when $\omega$ assumes a particular value at which
the curves are given by infinitely extended boundary lines that
divide the lens plane.

\input f5.cap

\subsection{square lattice}

For the square grid case, if $\omega<(\omega_0/\sqrt2)\approx1.311$
[i.e, $\kappa_\star>\pi/(2\omega_0^2)\approx0.457$], each of the critical
curves is centred at the centre of a square cell defined by the
four adjacent lenses (i.e., $k\omega+\ij p\omega$ where $k$ and $p$ are
odd integers). Any image in the region `within' the critical curves
has positive parity and vice versa. On the other hand, if
$\omega>2^{-1/2}\omega_0$ [$\kappa_\star<\pi/(2\omega_0^2)$], the critical
curves are centred around each lens location. For this case, the
images `within' the critical curves now have negative parity. Finally,
if $\omega=2^{-1/2}\omega_0$ [$\kappa_\star=\pi/(2\omega_0^2)$], the
critical curves are given by two infinite sets of parallel diagonal
lines, and the whole lens plane is evenly divided into checker-like
tiled regions according to the parity of the images in them.

For $\omega\ne2^{-1/2}\omega_0$, the lens equation maps each connected
and closed `unit' critical curve to a caustic that is also connected
and closed. Fig.~\ref{fig1} shows the critical curves and the
caustic network for $\omega=1.4$. Like the critical curves, the whole
caustic network is composed of the lattice arrangement of `unit'
caustic curves, but the unit curve exhibits self-crossing in contrast
to the critical curves. The shape of the unit caustic curve, which is
roughly similar to a variant of (irregular) octagram
(i.e., \{8/3\}-star-polygon\footnote{\scriptsize 
A $\{p/q\}$-star-polygon, with $p$, $q$ positive integers, is a figure
formed by connecting with straight lines every $q$-th point out of regularly
spaced $p$ points lying on a circumference. See
\url{http://mathworld.wolfram.com/StarPolygon.html}.}), is
actually generic for the case that $\omega>2^{-1/2}\omega_0$. The
`centre' of each unit caustic curve forms a similar square grid to the
lens lattice but the side length of the grid on the source plane is
given by $2\omega(1-\kappa_\star)$. As $\omega$ gets larger (or
equivalently $\kappa_\star$ gets smaller), the size of the caustics
shrinks and they asymptotically reduce to degenerate points. For
$\omega<2^{-1/2}\omega_0$, the caustic network can still be understood
as the lattice arrangement of `unit' caustic curves, the shape of
which is simply described as being a diagonally stretched
square. Again, the separation between the centres of the unit caustics
are given by $2\omega\abs{1-\kappa_\star}$. However, this is smaller than
the `size' of each unit caustic if $\kappa_\star\sim1$ so that the network
can exhibit overlapping among the neighboring caustics, which leads to
a rather complex network (albeit regular thanks to the symmetry of the
system). The critical curves and caustic network for $\omega=1.2$ are
shown in Fig.~\ref{fig2}. The transition of the caustic network over
$\omega=2^{-1/2}\omega_0$ can be understood as the contact between
(the vertices of) adjacent quasi-octagram caustics leading to their
connection. The further evolution of the caustic topography for
$\omega<2^{-1/2}\omega_0$ is presented in Fig.~\ref{fig3}. Since the
separation between the unit caustics reduces to nil as
$\kappa_\star\rightarrow1$, the network is the densest around
$\kappa_\star\approx1$. On the other hand, as $\kappa_\star$ increases past
unity, the separations between unit curves
$2\omega(\kappa_\star-1)$ increase while their sizes continually
decrease. Consequently, the network eventually reduces an array of
separate tiles of the unit caustics, which exhibit no overlap
between neighboring caustics. The critical value at which the sizes of
the unit caustics coincide with the separations between them is
found approximately to be $\omega\approx0.748$ ($\kappa_\star\approx1.405$).

\input f6.cap

One physical interpretation of the caustics in gravitational
lensing is that they are boundaries between regions in the source
plane that produce different number of images. In addition, it is also
known that a pair of images of opposite parity appears or disappears
whenever the source crosses the caustics, and that the number of
positive parity images, if the source lies outside any caustics, is
one or nil depending on the characteristics of the system. Hence, the
maximum number of positive parity images can be determined from
examination of the caustics network. For example, the
quasi-octagram caustics for $\omega>2^{-1/2}\omega_0$ allow at most
four positive parity images in the region around the centres of
caustics. It is easy to convince oneself that the minimum number of
positive parity images for the corresponding scenario is one when
the source lies outside any caustic. As for the case
$\omega\le2^{-1/2}\omega_0$, the examination of the caustic network
leads us to conclude that, while the maximum number of the positive
images approaches infinity at $\kappa_\star=1$ (i.e.,
$\omega=\pi^{1/2}/2\approx0.886$), it is greater than four only if
$0.803\la\omega\la1.039$. In other words, the number of positive
parity images is bounded by four provided that $\omega\ga1.039$
($\kappa_\star\la0.727$) or that $\omega\la0.803$ ($\kappa_\star\ga1.217$).
The source positions with no positive parity image start to exist if
$\omega\la0.776$ or equivalently $\kappa_\star\ga1.039$. Finally, if
$\omega\la0.748$ ($\kappa_\star\ga1.405$), there is no overlap between
neighboring caustics, and thus the number of positive parity
images is one if the source lies in the caustic, and nil if otherwise.

\subsection{equilateral triangular lattice}

\input f7.cap

The lenses on an equilateral triangular lattice produce the critical
curves and the caustic network in a basically consistent pattern as
the square lattice lens although many details are quite different.
For triangular lattices, the unit critical curves for
$\omega<(\omega_2/\sqrt[3]2)\approx1.214$
[$\kappa_\star>\pi/(2^{1/3}3^{1/2}\omega_2^2)\approx0.615$] are located in
each of the equilateral triangular cell defined by three adjacent lenses.
On the other hand, those for $\omega>2^{-1/3}\omega_2$
[$\kappa_\star>\pi/(2^{1/3}3^{1/2}\omega_2^2)$] are centred at each lens
position. Note that, in contrast to the square lattice, the number of
positions with zero shear is twice that of the poles (i.e., the lens
positions) so that there is two-to-one correspondence between the unit
critical curves of the former and that of the latter. At the critical
value that $\omega=2^{-1/3}\omega_2$
[$\kappa_\star=\pi/(2^{1/3}3^{1/2}\omega_2^2)$], the lens plane is divided
into two classes by the critical curves: triangular regions around the
null shear positions within which the images have positive parity
and hexagonal regions around the lenses corresponding to the images of
negative parity.

For $\omega>2^{-1/3}\omega_2$, the shape of the unit caustics may be
described as snowflake-like or an irregular quasi-dodecagram
(\{12/5\}-star-polygon). Each unit caustic is arranged in the same
equilateral triangular grid with side length of
$2\omega(1-\kappa_\star)$ to form the whole caustic network.
Fig.~\ref{fig1t} shows an example of this, the caustic network for
$\omega=1.25$. The maximum number of positive parity images associated
with these unit caustics is found to be six. If
$\omega<2^{-1/3}\omega_2$, the unit caustics are given by stretched
triangular cells, and the whole network is given by tiling the
source plane with these cells, which may overlap with one another. Here,
the centres of each cell actually form a hexagonal (honeycomb) grid
with side length of $2\omega\abs{1-\kappa_\star}/\sqrt3$, which is in
fact the mapped image of the dual grid to the lens lattice in the lens
plane, and the adjacent cells alternate their orientation with their
symmetry maintained. An example of these caustics networks is
presented in Fig.~\ref{fig2t} for $\omega=1.15$. Analogous to the
case of the square lattice, dense overlapping of unit caustics leads to
very complex network around $\kappa_\star\sim 1$. It is found that that
the critical value that corresponds to no overlap between adjacent
caustics is about $\omega\approx0.791$ ($\kappa_\star\approx1.449$). The
triangular lens lattice with side length less than this value
(hence a denser lattice) produces triangular unit caustics arranged on
a honeycomb grid with no overlap, and the number of positive
parity images is either one or nil depending on the position of the
source relative to the caustics.

\input f8.cap

\subsection{regular hexagonal lattice}

In qualitative terms, the critical curves of the honeycomb lattice are
basically the `dual' of that of the triangular case. The critical
curves for $\omega<(2^{2/3}3^{-1/2}\omega_2)\approx1.402$
[$\kappa_\star>\pi/(3^{1/2}2^{4/3}\omega_2^2)\approx0.308$] are analogous
to those for the triangular lattice with $\omega>2^{-1/3}\omega_2$ and
located within each hexagonal cell centred at the positions of null
shear. On the other hand, if $\omega>2^{2/3}3^{-1/2}\omega_2$, the
critical curves are like those for the triangular lattice with
$\omega<2^{-1/3}\omega_2$ and centred around each lens positions. The
dual characteristic further indicates that there are twice more poles
than the locations of null shear (which are in fact fourth-order
zeros), and thus there is one-to-two correspondence between unit
critical curves of the former to the latter (formally, the `winding
number' of the former is twice that of the latter).

On the other hand, the caustics of a honeycomb lattice are
qualitatively distinct from those of a triangular lattice.
Fig.~\ref{fig1h} shows an example for a sparse honeycomb lattice with
$\omega=1.45$, which is in fact archetypal for
$\omega>2^{2/3}3^{-1/2}\omega_2$. The unit caustics are given by a
six-sided closed self-intersecting curve, which may be labeled a
bi-triangle, and they are arranged in a similar hexagonal pattern as
the lattice of the lens with side length $2\omega(1-\kappa_\star)$.
We also note that the maximum number of positive parity images
associated with these caustics is three. For a dense honeycomb lens
lattice ($\omega<2^{2/3}3^{-1/2}\omega_2$), the unit caustics are
found to be in the shape of a stretched hexagon (an example of which
for $\omega=1.3$ is given in Fig.~\ref{fig2h}) and arranged in a
triangular grid with side length of $2\sqrt3\omega\abs{1-\kappa_\star}$.
They also exhibit trends of varying overlap similar to the square or
triangular lattice, as $\kappa_\star$ gets larger past the unity. We find
that the critical value for no overlapping caustics for the honeycomb
lattice is approximately given by $\omega\approx0.622$
($\kappa_\star\approx1.565$).

Despite their variety, we also notice some common characteristics of
the caustic networks of three kinds of regular lensing lattices.
For example, with an optical depth that is greater than a certain
critical value, the unit curves are more or less in a similar shape
as the lensing lattice except for the fact that they are distorted
in a way stretching the vetices radially outward. They densely overlap
with one another around $\kappa_\star\sim1$. However, as $\kappa_\star$ gets larger
past the unity, their size shrink whereas the separations among them grow,
and so the networks eventually reduce to a simple array of tiles.
Likewise, towards the lower optical depths below the critical value,
the networks are comprised of star-shaped unit curves whose
exact characteristics are related to meeting pattern of 
the lensing lattice at the common vertex points.

\section{Mean number of positive parity images}
\label{sec:mn}

\begin{figure}
\includegraphics[width=\hsize]{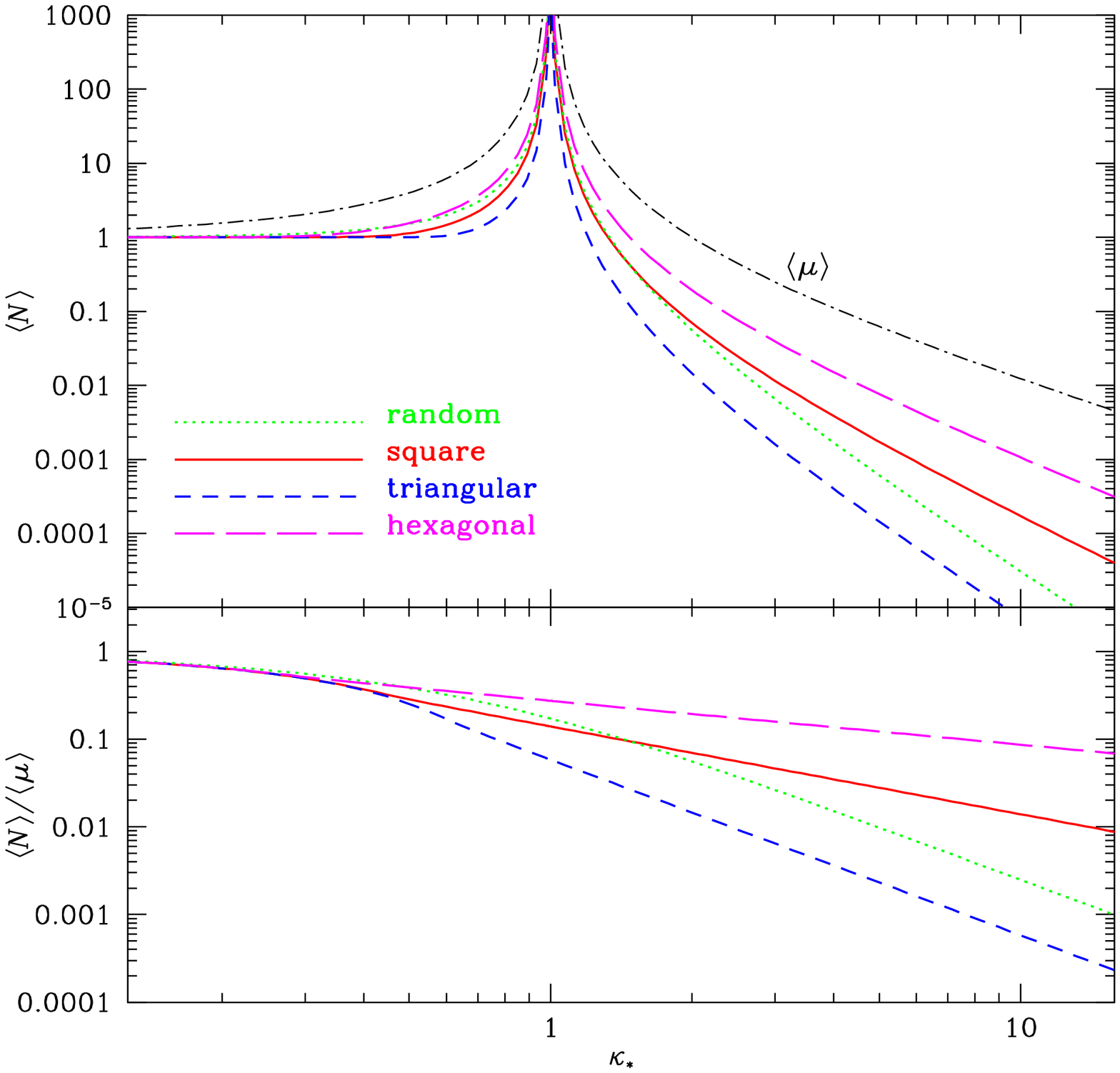}
\caption{\label{fig4} Top: mean number of positive parity image for
lenses on a square lattice (solid line), a triangular lattice
(short-dashed line), a hexagonal lattice (long-dashed line), and for
the random star field (dotted line). Also plotted in dot-dashed lines
is the mean total magnification $\langle\mu\rangle=(1-\kappa_\star)^{-2}$.
Bottom: mean number normalized by the mean total magnification. The
line types are the same.}
\end{figure}

The area in the caustic accounting for its multiplicity
can be understood as the covering fraction of the source plane by the
region with `extra image' pairs once it is properly normalized. In
general, however, the normalization is ill-defined for localized
lenses. This difficulty is ameliorated with the introduction of the
lattice of point masses since the whole plane is now evenly divided
into identical cells. With the regular lensing lattice, the area under
the unit caustic can be properly normalized with respect to the area
of the mapped image (on the source plane) of the unit cell comprising
the lensing lattice. The result, if the multiplicity is properly
accounted for, is actually the mean number of extra image pairs, which
has been calculated for the case of the random star field using
techniques from the probability theory \citep{WWS,GSW}.

\input f10.cap

For `dense' lattices that produce nonintersecting (but possibly
overlapping) unit caustics, the relevant area under the unit caustics
is straightforward to calculate since, despite overlapping, they are
simply reduced to a set of closed loops. Then, the area under one such
loop normalized to the area of a unit cell in the source plane,
$\mathscr A_\omega(1-\kappa_\star)^2$ where $\mathscr A_\omega$ is the
area under the unit cell defined by the adjacent lens positions in the
lens plane, is the mean number both of extra image pairs and of
positive parity images for a given source position because the
corresponding critical curves in the image plane completely enclose
the region where any positive images reside and thus there are no
positive images formed `outside' of any caustics. On the other hand,
for `sparse' lattices that produce the network composed of a grid set
of self-intersecting `star-shaped' curves, the self-crossing of the
unit structures requires some consideration on the meaning of `area'
that properly accounts for the image multiplicity. We argue that the
solution is rather simple. Once the area calculation is reduced to the
line integral along its boundary \citep[see e.g.,][]{An05,AE06} using the
fundamental theorem of multivariate calculus\footnote{\scriptsize 
The theorem is usually
known as Green's theorem in elementary multivariate calculus
when referring to the relation between two-dimensional integral
over a domain in two-dimensional space and one-dimensional integral
along the boundary of the domain. This is also a special case of
so-called (generalized) Stoke's theorem, the name under which the theorem
is usually known as.},
the corresponding line integral along the caustic actually results in
the area counted with multiplicity. That is to say,
they can be considered as the
`area-excess' or the `over-covering factor' of the source positions
that produce positive parity images. Unlike the previous case, the
normalization cell in the lens plane is the unit cell of the dual
of the lensing lattice, i.e., the cell defined by the adjacent null
shear locations (which is in fact the centre of each cell of the
lensing lattice). The corresponding critical curves lie completely
within this unit normalization cell and furthermore enclose the
regions of the image plane where negative parity images reside.
Subsequently, there is at least one positive parity image for any
given source position -- that is, the source `outside' caustics forms
one positive parity image. As a result, the mean number of extra image
pairs is less than the number of positive parity images by one. We
note that the mean number of positive parity images actually varies
continuously over the critical value dividing the dense lattice from
the sparse one while the mean number of extra image pairs jumps by
one. In fact, designating a certain image pair as an extra involves
some degrees of arbitrariness whereas counting the number of positive
parity images is well-defined. For these reasons, henceforth, we shall
discuss the mean number of positive parity images exclusively.

Fig.~\ref{fig4} shows the resulting mean number of positive parity
images as a function of $\kappa_\star$. Comparison to the case of the
random star field (with zero external shear) \citep{WWS} reveals some
significant differences between the regular lensing lattices and the
random field, particularly in the behaviour toward
$\kappa_\star\rightarrow\infty$. While the mean number falls off as
$\kappa_\star^{-4}$ for the random star field and triangular lensing
lattices (albeit the normalization for the random field being larger
by a factor of $\sim$4), its behaviours for the lenses on square
lattices and honeycomb lattices are characterized by slower fall-offs
given by $\sim\kappa_\star^{-3}$ and $\sim\kappa_\star^{-2.5}$, respectively.
At first, this is somewhat counterintuitive. That is, in the random
star field, the distance $s$ to the closest neighboring lens is
distributed according to
\begin{equation}
\mathscr P(s)\,\df s=2\pi\sigma s\e^{-\pi\sigma s^2}\,\df s
\end{equation}
where $\sigma=\kappa_\star/\pi$ is the mean surface number density of the
stars. The mean and the variance are given by
$\omega=1/(2\sqrt\sigma)$ and
$[(4/\pi)-1]\omega^2=[1-(\pi/4)]\kappa_\star^{-1}$. In other words, as the
density of stellar field increases, the random fluctuations around the
mean decrease and one would expect that the random stellar field
may approach to a regular field.

The actual result can be understood using the distribution of the
(internal) lensing shear $\gamma$ due to stars experienced by the given line of
sight towards a lensed image. For a field of point masses, once this
distribution is known, the mean number of positive parity images
is recovered through \citep{WWS}
\begin{equation}
\label{eq:intg}
\langle N_+\rangle=
\langle\mu\rangle\int_0^1\!(1-\gamma^2)\mathscr P(\gamma)\,\df\gamma.
\end{equation}
With lenses on a regular lattice, we have shown that the shear at the
given image plane location is given by the Weierstrass elliptic
function, and hence, the distribution on the whole image plane reduces
to that on the unit cell in the image plane. Then, the cumulative
distribution of shear is found by the fraction of area in the unit
cell where the shear is smaller than the given value, and
consequently, its derivative gives the differential distribution of
the shear. Furthermore, the homogeneity relation of $\wp$-function
implies the similarity relation between the distribution corresponding
to different $\omega$. For example, with a square lensing lattice,
we have that
\begin{gather}
\begin{split}
\int_0^\gamma\mathscr P(\gamma'\vert\kappa_\star)\,\df\gamma'&
=\frac1{\mathscr A_\omega}
\mathscr A_\omega\Bigl[\abs{\wp(z;g_2,0)}\le\gamma\Bigr]
\\&
=\frac1{\mathscr A_{\omega_0}}
\mathscr A_{\omega_0}\!\Biggl[
\abs{\wp\Bigl(\frac\omega{\omega_0}\,z;1,0\Bigr)}
\le\frac{\omega^2}{\omega_0^2}\,\gamma
=\frac{\pi}{4\omega_0^2}\frac{\gamma}{\kappa_\star}\Biggr]
\\&
=\int_0^{\kappa_0\gamma/\kappa_\star}
\mathscr P(\gamma'\vert\kappa_0)\,\df\gamma'
\end{split}
\\
%
\label{eq:gdis}
\begin{split}
\mathscr P(\gamma\vert\kappa_\star)&
=\frac1{\mathscr A_{\omega_0}}
\Biggl.\frac\df{\df\gamma'}\mathscr A_{\omega_0}\biggl[
\Bigl\lvert\wp\Bigl(\frac\omega{\omega_0}\,z;1,0\Bigr)\Bigr\rvert
\le\frac{\kappa_0}{\kappa_\star}\,\gamma'\biggr]\Biggr|_{\gamma'=\gamma}
\\&
=\frac1{\mathscr A_{\omega_0}}\frac{\df g}{\df\gamma}
\Biggl.\frac\df{\df g}\mathscr A_{\omega_0}\!\biggl[
\Bigl\lvert\wp\Bigl(\frac\omega{\omega_0}\,z;1,0\Bigr)\Bigr\rvert
\le g\biggr]\Biggr|_{g=\kappa_0\gamma/\kappa_\star}
\\&
=\frac{\kappa_0}{\kappa_\star}\
\mathscr P\biggl(\frac{\kappa_0}{\kappa_\star}\,\gamma\,\vrule\,\kappa_0\biggr)
\end{split}
\end{gather}
where $\kappa_0=\pi/(2\omega_0)^2$ and $\mathscr A_\omega
[\abs{\wp(z;g_2,0)}\le\gamma]$ is the area of the unit cell where
$\abs{\wp(z;g_2,0)}\le\gamma$ for a square lensing lattice with
side length of $\omega$ (see eq.~[\ref{eq:g2}] for $g_2$ as a function
of $\omega$ and $\kappa_\star$). In fact, any value of $\kappa_\star$ instead
of $\kappa_0$ can be chosen as the normalization and the resulting
similarity relation is valid for all three regular lattices.

\begin{figure*}
\includegraphics[width=0.8\hsize]{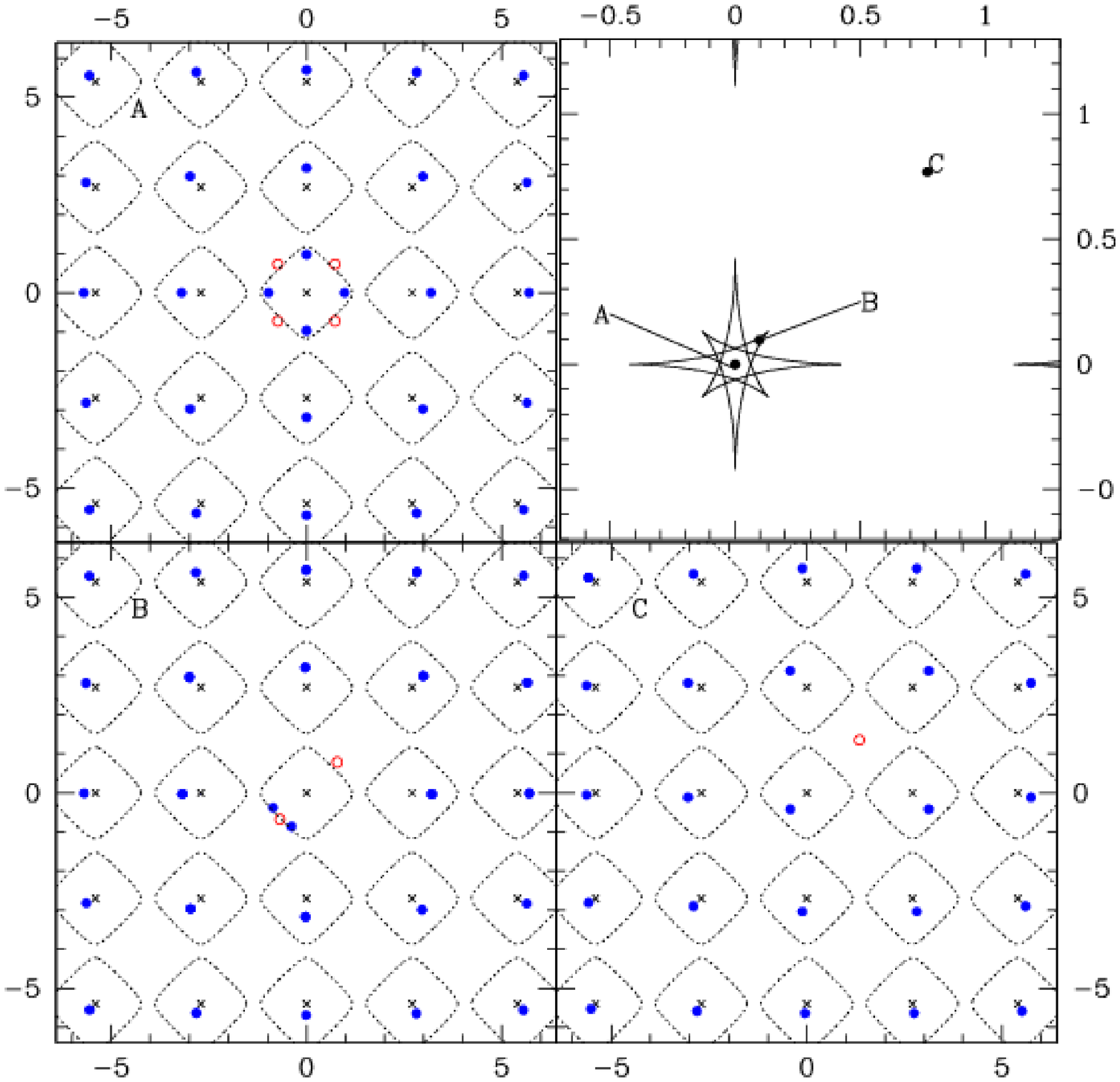}
\caption{\label{figim1} Locations of lensed images with a square lensing
lattice with $\omega=1.35$ ($\kappa_\star\approx0.431$).
The locations of the source with respect
to the caustic networks are indicated in the top right panel. In the
remaining three panels, the locations of the individual lensed images
are marked by open circles (positive parity images) or closed dots
(negative parity images). Also drawn in dotted lines are the critical curves
and the locations of the lenses are marked by crosses.}
\end{figure*}

\begin{figure*}
\includegraphics[width=0.8\hsize]{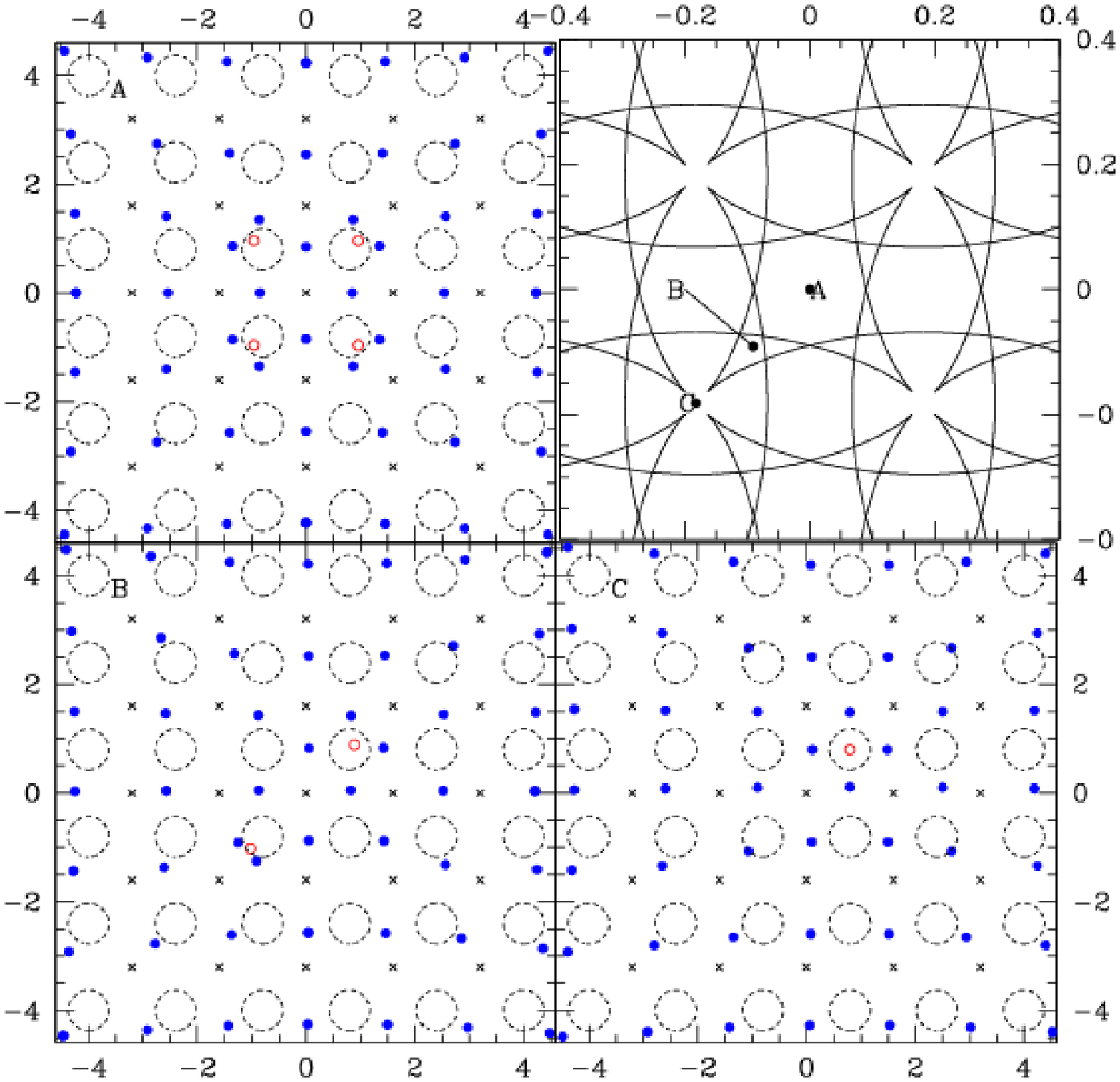}
\caption{\label{figim2} Same as Fig.~\ref{figim1} except $\omega=0.8$
($\kappa_\star\approx1.23$).}
\end{figure*}

\begin{figure*}
\includegraphics[width=0.8\hsize]{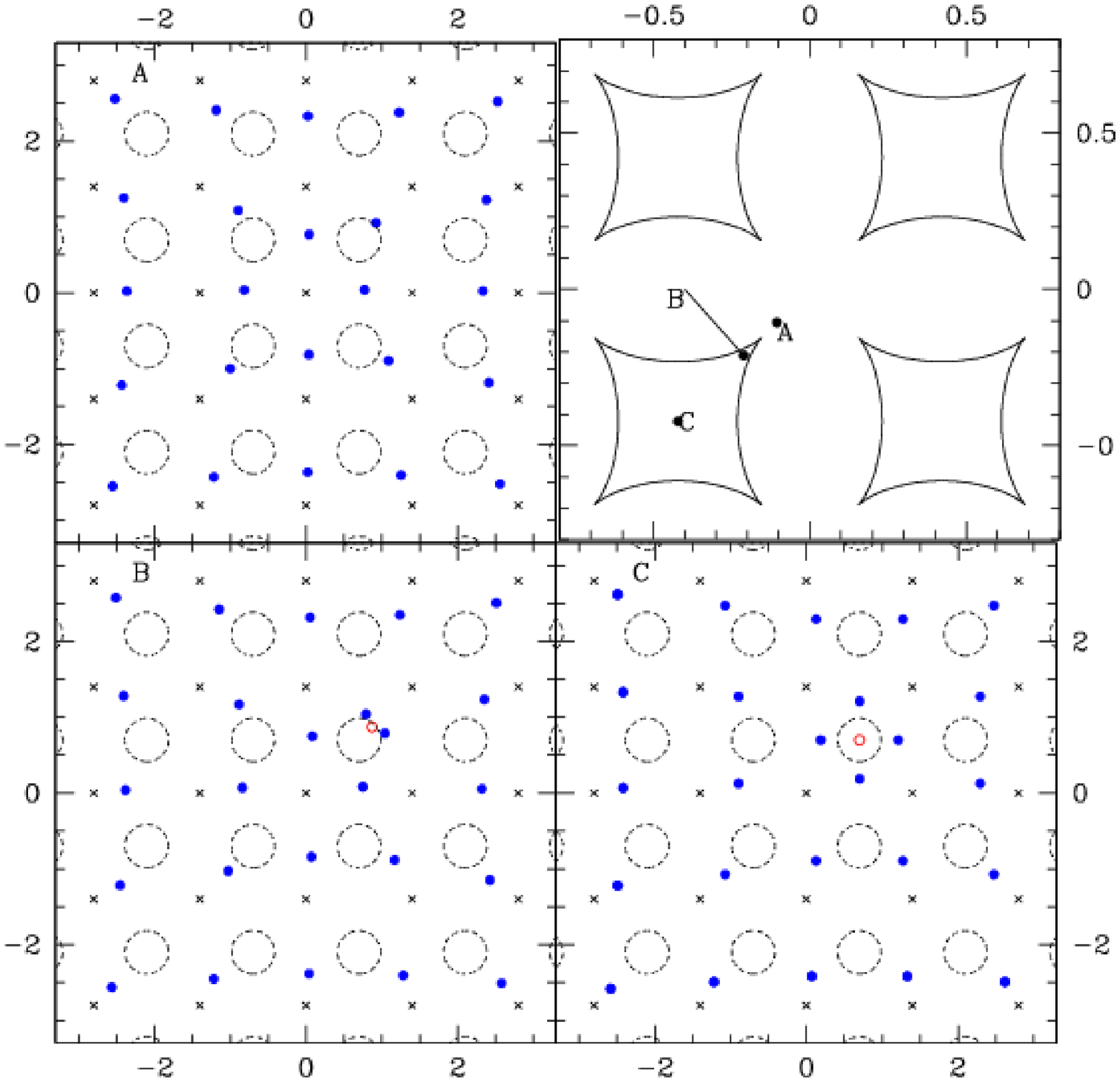}
\caption{\label{figim3} Same as Fig.~\ref{figim1} except $\omega=0.7$
($\kappa_\star\approx1.60$). Shown in the top left panel, there is no
positive parity images for the source lying `outside' the caustics.}
\end{figure*}

\begin{figure*}
\includegraphics[width=0.8\hsize]{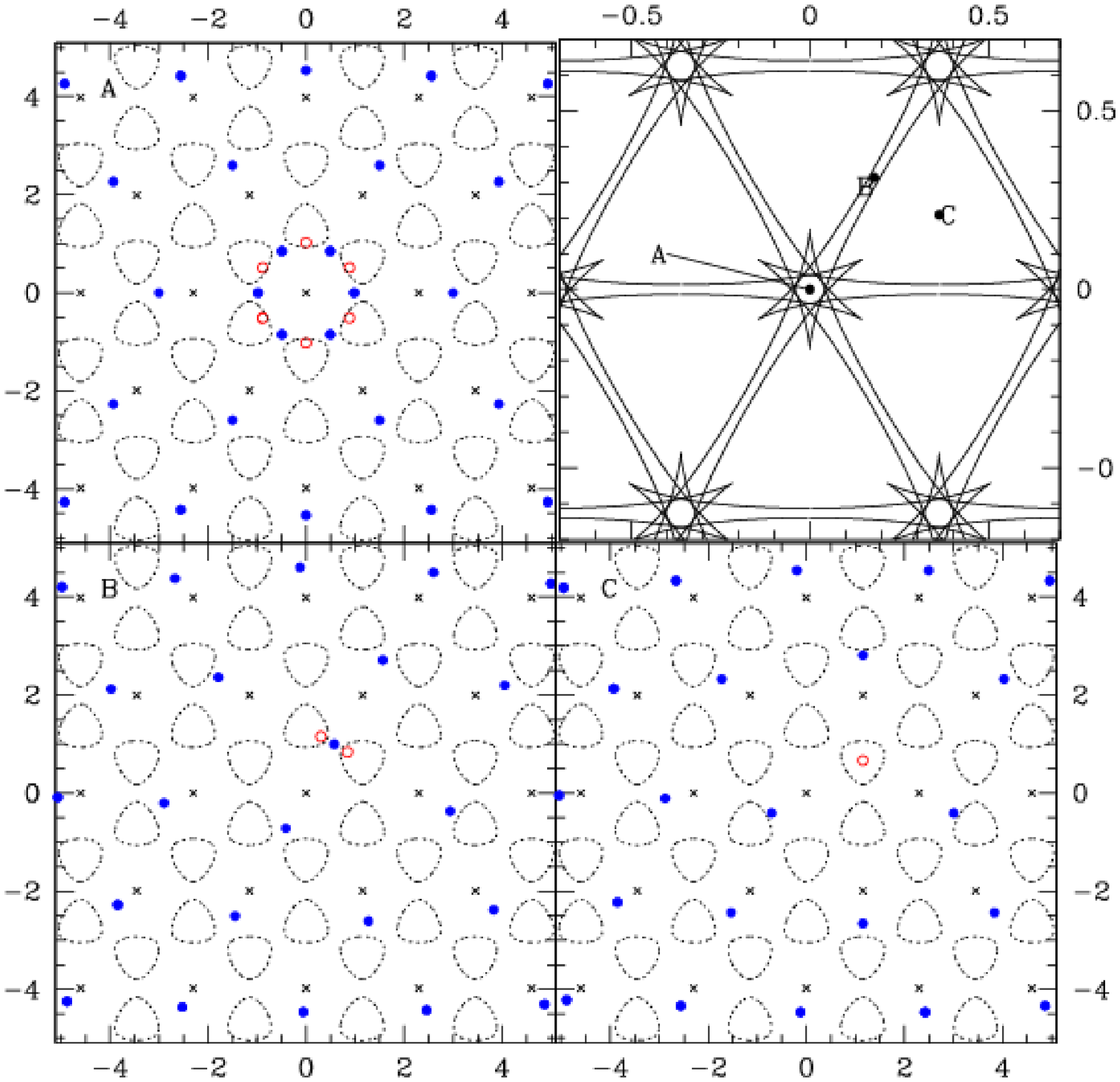}
\caption{\label{figimt} Same as Fig.~\ref{figim1} except for an equilateral
triangular lensing lattice with $\omega=1.15$ ($\kappa_\star\approx0.685$).}
\end{figure*}

The resulting distributions
for all three regular lattices are shown in Fig.~\ref{fig5} together
with the same distribution for the random star field
\citep{NO84,Sc87,WWS}, which is given by
\begin{equation}
\mathscr P_\mathrm{ran}(\gamma\vert\kappa_\star)
=\frac{\kappa_\star\gamma}{(\kappa_\star^2+\gamma^2)^{3/2}}
=\frac1{\kappa_\star^2}\frac\gamma{\left[1+(\gamma/\kappa_\star)^2\right]^{3/2}}.
\end{equation}
Unlike the random field case, we find that there is nonzero
probability that the line of sight experiences no net shear for the
square lattice lens. By contrast, the same distribution linearly tends
to zero as $\gamma\rightarrow0$ for the triangular lattice (and also
the random field) and it diverges as $\sim\gamma^{-1/2}$ for the
honeycomb lattice. These behaviours are related to the fact that the zeros
of $\wp(z;g_2,0)$, $\wp(z;0,g_3)$ and $\mathfrak h(z;g_3)$ are
second, first, and fourth-order, respectively; that is to say,
their Taylor-series expansions at their zeroes are given by
\begin{equation}
\begin{split}&
\wp(z;g_2,0)\simeq-\frac{g_2}4\,(z-z_0)^2+\mathscr O(\abs{z-z_0}^6)
\\&
\wp(z;0,g_3)\simeq\ij g_3^{1/2}\,(z-z_0)+\mathscr O(\abs{z-z_0}^4)
\\&
\mathfrak h(z;g_3)\simeq\frac{g_3}{27}\,z^4+\mathscr O(z^{10})
\end{split}
\end{equation}
%
where $z_0$ is a zero of the corresponding function (note that zero of
$\mathfrak h$ is at $z_0=0$). For $\abs{\gamma}\ll1$, the discussion
in the preceding paragraph indicates that
$\mathscr A_\omega\int_0^\gamma\mathscr P(\gamma')\df\gamma'\simeq
\pi\abs{z-z_0}^2$ where $z_0$ is the null shear position and $z$
traces the locations at which the shear is given by $\gamma$. By
noticing the shear is given by $\abs{\wp(z;g_2,0)}$,
$\abs{\wp(z;0,g_3)}$ and $\abs{\mathfrak h(z;g_3)}$ for square,
triangular, and honeycomb lattices, we can then find the leading-order
behaviours for the shear distributions as $\gamma\rightarrow0$ from the
above Taylor expansions\footnote{\scriptsize It is actually easier to work out
from the series for the inverse functions, which are in fact
hypergeometric function (see also Appendix).};
\begin{equation}
\begin{split}
\mathscr P_{\ssqu}(\gamma\vert\kappa_\star)&
\simeq\frac1{\kappa_\star}\,\frac{2^6\pi^4}{\Gamma^8_{1/4}}+
\mathscr O\biggl[\Bigl(\frac\gamma{\kappa_\star}\Bigr)^2\biggr]
\approx\frac{0.2088}{\kappa_\star}
\\
\mathscr P_{\stri}(\gamma\vert\kappa_\star)&
\simeq\frac1{\kappa_\star}\,\frac{2^{11}\pi^9}{3^{3/2}\Gamma^{18}_{1/3}}\,
\frac\gamma{\kappa_\star}+
\mathscr O\biggl[\Bigl(\frac\gamma{\kappa_\star}\Bigr)^4\biggr]
\approx\frac{0.2326}{\kappa_\star^2}\,\gamma
\\
\mathscr P_{\shex}(\gamma\vert\kappa_\star)&
\simeq\frac1{\kappa_\star}\,\frac{2^4\pi^{9/2}}{3^{3/4}\Gamma^9_{1/3}}\,
\left(\frac{\kappa_\star}\gamma\right)^{1/2}
+\mathscr O\Bigl(\frac\gamma{\kappa_\star}\Bigr)
\approx\frac{0.1705}{\kappa_\star^{1/2}\gamma^{1/2}},
\end{split}
\end{equation}
which agree with the numerical results shown in Fig.~\ref{fig5}
(note that $\kappa_\star=\pi$ for Fig.~\ref{fig5}). In other words, the
higher-than-linear-order (i.e., degenerate) zeros of the shear lead to
nonzero finite or even divergent probability for the zero net shear.
Physically speaking, accidental canceling of the net shear in the
random star field is much less likely to occur than in the highly
symmetric arrangement of the lenses, for which there may be
significant chances that the image is located at an exactly balanced
position where the net shear is null. This is basically the reason
for slower fall-offs of the mean number of positive parity images in
dense square or honeycomb lattices. From equation (\ref{eq:gdis}), if
the shear distribution behaves like $\mathscr P(\gamma\vert\kappa_\star)
\simeq S\kappa_\star^{-1}(\gamma/\kappa_\star)^\alpha$ as
$\gamma/\kappa_\star\rightarrow0$ where $S$ is a constant,
$\alpha=(2/n)-1>-1$ is the asymptotic power index for the shear
distribution and $n$ is the order of the zeros of the shear, then
equation (\ref{eq:intg}) as $\kappa_\star\rightarrow\infty$ reduces to
\begin{equation}
\begin{split}
\langle N_+\rangle&
=\frac1{(1-\kappa_\star)^2}
\int_0^1\!\df\gamma\,(1-\gamma^2)\mathscr P(\gamma\vert\kappa_\star)
\\&
\simeq\frac S{(\kappa_\star-1)^2\kappa_\star^{\alpha+1}}
\int_0^1\!\df\gamma\,(1-\gamma^2)\gamma^\alpha.
\end{split}
\end{equation}
Since the last integral is a finite constant for $\alpha>-1$, we find
that $\langle N_+\rangle\sim\kappa_\star^{-(\alpha+3)}$ as
$\kappa_\star\rightarrow\infty$, which is consistent with the results
found earlier (Fig.~\ref{fig4}).

In fact, the shear distributions for regular lensing lattices are
quite distinct from that of the random field, and it is somewhat
remarkable that the behaviours of $\langle N_+\rangle$ are generally
similar for all cases at least for small $\kappa_\star$. This is related
to the fact that $\langle N_+\rangle$ is basically an integration (or
convolution) of the shear distribution, and therefore any `roughness'
in the shear distribution is `smoothed' out in $\langle N_+\rangle$.
In particular, we note that the similarity relation of the shear
distribution is essentially the result of the scale invariance and so
should be generic. Then, $\mathscr P(\gamma\vert\kappa_\star)=
\kappa_\star^{-1}\mathcal P(\gamma/\kappa_\star)$ where
$\mathcal P(\gamma/\kappa_\star)$ is a function of $\gamma/\kappa_\star$, and
consequently we have
\begin{equation}
\langle N_+\rangle=\frac1{(1-\kappa_\star)^2}
\int_0^{1/\kappa_\star}\!\df t\,(1-\kappa_\star^2t^2)\,\mathcal P(t).
\end{equation}
Furthermore, since the shear for the lines of sight that pass
sufficiently close to any of the lenses is simply given by
$\gamma\simeq d^{-2}$ where $d$ is the separation between the given
line of sight and that to the lens, it is straightforward to establish
that the generic asymptotic behaviour of the shear distribution as
$\gamma/\kappa_\star\rightarrow\infty$ to be
$\mathscr P(\gamma\vert\kappa_\star)\simeq\kappa_\star/\gamma^2$ [i.e.,
$\mathcal P(t)\simeq t^{-2}$ as $t\rightarrow\infty$]. Suppose that
power-series expansion of $\mathcal P(t)$ at $t=\infty$ is given by
$\mathcal P(t)\simeq t^{-2}(1+\sum_{n=1}^\infty c_nt^{-pn})$ (and it
is uniformly convergent in an interval containing $t=\infty$). Then,
for $\kappa_\star\rightarrow0$, we find that (assuming $p\ne1/n$ for any
positive integer $n$);
\begin{equation}
\begin{split}&
\int_0^{1/\kappa_\star}\!\df t\,\mathcal P(t)\simeq
1-\kappa_\star-\sum_{n=1}^\infty\frac{c_n\kappa_\star^{pn+1}}{pn+1}
\\&
\int_0^{1/\kappa_\star}\!\df t\,t^2\,\mathcal P(t)\simeq
\frac1{\kappa_\star}-P_0-\sum_{n=1}^\infty\frac{c_n\kappa_\star^{pn-1}}{pn-1}
\\&
\frac{\langle N_+\rangle}{\langle\mu\rangle}\simeq
1-2\kappa_\star+P_0\kappa_\star^2
+\sum_{n=1}^\infty\frac{2c_n\kappa_\star^{pn+1}}{(pn-1)(pn+1)}
\end{split}
\end{equation}
where $P_0$ is a finite constant that weakly depends on the global
(that is, far from $t=\infty$) behaviour of $\mathcal P(t)$, and thus
$\langle N_+\rangle/\langle\mu\rangle\simeq1-2\kappa_\star+
\mathscr O(\kappa_\star^2)$ and
$\langle N_+\rangle=1+\mathscr O(\kappa_\star^2)$,
regardless of the higher-order behaviour of $\mathcal P(t)$, provided
that $p>1$. In fact, we have $\mathcal P_\mathrm{ran}(t)=
t(1+t^2)^{-3/2}=t^{-2}(1+t^{-2})^{-3/2}$ so that $p=2$ for
$\mathcal P_\mathrm{ran}(t)$, and also the power-series expansions of
$\wp$-functions at their poles and zeros (or equivalently the
hypergeometric function form of their inverse functions) imply that
%
the power-series expansions of $\mathcal P(t)$ at $t=\infty$ for
regular lensing lattices are given in the above form with $p=2$,
$p=3$, and $p=3/2$ for the square, triangular, and honeycomb lattices,
respectively.
Therefore, all four cases exhibit the common asymptotic behaviour of
$\langle N_+\rangle$ for $\kappa_\star\ll1$.

\subsection{image locations}

While the overall characteristics of the lensing situation can be
understood simply by studying the critical curves and caustics, one
still needs to invert the lens equation to find the image locations
for a given source position. With transcendental functions involved,
this must be done numerically, but some analytical insights can ease
the task.

Since the real axis of the system for the lens equations
(\ref{eq:le}), (\ref{eq:leq_tri}), and (\ref{eq:leq_hex}) is chosen
such that the lensing lattice exhibits reflection symmetry with
respect to this axis, the deflection functions for all cases should
behave $\overline{f(z)}=f(\z)$, which is indeed confirmed by the
symmetry of the Weierstrass functions. Then, the lens equations
(\ref{eq:le}), (\ref{eq:leq_tri}), and (\ref{eq:leq_hex}) can also be
written to be
\begin{equation}
\label{eq:lea}
\eta=z-f(\z)\,;\qquad
f(z)=
\begin{cases}
\zeta(z;g_2,0)&\squ\\
\zeta(z;0,g_3)&\tri\\
\mathfrak d(z;g_3)&\hex
\end{cases}.
\end{equation}
Applying (two-dimensional) Newton-Raphson method to equation (\ref{eq:lea})
indicates that the solutions are given by the $(n\rightarrow\infty)$-limits
of the sequences given by the recursion relation
\begin{equation}
\label{eq:sec1}
z_{n+1}
=\frac{\eta+f(\z_n)-[\z_n-\bar\eta-f(z_n)+z_nf'(z_n)]f'(\z_n)}
{1-f'(\z_n)f'(z_n)}.
\end{equation}
Alternatively, we can eliminate $\z$ from equation (\ref{eq:lea}) by means of
their complex conjugate \citep[e.g.,][]{WM95,AE06}
\begin{equation}
\label{eq:lec}
\z=\bar\eta+f(z).
\end{equation}
and consequently, the image positions corresponding to the source
position of $\eta$ are the fixed points of the `conformal' mapping
given by
\begin{equation}
h(z)=\eta+f\bigl[\bar\eta+f(z)\bigr]\,;\qquad
h'(z)=f'\bigl[\bar\eta+f(z)\bigr]\,f'(z)
\label{eq:map}
\end{equation}
where
\begin{equation}
f'(z)=
\begin{cases}
-\wp(z;g_2,0)&\squ\\
-\wp(z;0,g_3)&\tri\\
-\mathfrak h(z;g_3)&\hex
\end{cases}.
\label{eq:fp}
\end{equation}
In principle, all image positions can again be found applying the (complex)
Newton-Raphson method to zeros of $z-h(z)$ (which include all image
positions and possibly some spurious solutions) unless the image is
right on the critical curve. Specifically, the zeros can be located
from the limits of the sequences given by
\begin{equation}
\label{eq:sec2}
z_{n+1}
=z_n-\frac{z_n-h(z_n)}{1-h'(z_n)}
=\frac{z_nh'(z_n)-h(z_n)}{h'(z_n)-1}
\end{equation}
as $n\rightarrow\infty$. While equation (\ref{eq:sec2}) is basically
the same as equation (\ref{eq:sec1}) except for the fact that
$\z_n$ in equation (\ref{eq:sec2}) is replaced by $\bar\eta+f(z_n)$
in equation (\ref{eq:sec1}), the latter relation in general is
slightly better-behaved and also converges faster than the former.
The iteration (with infinite precision) given by equation (\ref{eq:sec2})
necessarily converges to a zero starting from anywhere in the complex
plane except for a set of points with zero measure (`Julia set'), and
furthermore any solution that is not on the critical curve has a basin
of attraction that contains an open neighborhood of the
solution. However, where the iteration converges to for a given
starting point is nontrivial to figure out a priori.

In fact, if we want to locate all of positive parity images, there
actually exists an alternative route which is much more economical.
Application of the results from the complex dynamics \citep[see also
Appendix of][]{AE06} implies that, for any solution $z_0$ of equation
(\ref{eq:lec}) that is also $\abs{f'(z_0)}<1$, there exists a point
$z_c$ such that $f'(z_c)=0$ and
$\lim_{n\rightarrow\infty}h^n(z_c)=z_0$. Here, $h^1(z)=h(z)$ and
$h^n(z)=h(h^{n-1}(z))$, that is, $h^n(z)$ is the $n$-times iterations
of the mapping $h$ starting from $z$. In other words, all the positive
parity image positions are the limit points of the iterations of the
mapping $h(z)$ given by equation (\ref{eq:map}) starting from a
zero of $f'(z)$ (eq.~[\ref{eq:fp}]). Conversely, the iterations of the
mapping $h(z)$ starting from any zero of $f'(z)$ necessarily converge
to one of the fixed points of $h(z)$, the set of which contains all the
positive parity image positions. Since the number of positive parity
images can be determined by examining the caustic network,
one only needs, in principle, to perform the iterations finitely many
times until all images are located (unless $\kappa_\star=1$). In addition,
all the zeros of $f'(z)$ for regular lensing lattices occur at the
centre of each unit cell defined by the adjacent poles (here, the
poles are simply the lens positions). In summary, all the positive parity
image positions for a given source position can be located by
the iterations of $h(z)$ starting from the centres of
each unit cell (corresponding to the null shear positions) until the
set of distinct limit points contains the number of image positions
of positive parity that is predicted by the source
position and the caustic network.

As for negative parity images, although there are infinitely many
of them, we argue that most of them are insignificant. First, the lens
equations indicate that
$\abs{z-\eta}=\lvert\overline{f(z)}\rvert=\abs{f(z)}$, which implies
that, in order to have an image that is arbitrarily far from the
source, the image position should be rather close to the lens, that
is, the poles of $f(z)$, at which $f(z)$ is divergent. If we consider
the lens mapping from the point $z=\delta z+\Omega_{mn}$ where
$\Omega_{mn}$ is one of the lens locations, then the lens equation
(\ref{eq:lea}) reduces to
\begin{equation}
\label{eq:ape}
\eta=\delta z+\Omega_{mn}-f\bigl(\overline{\delta z+\Omega_{mn}}\bigr)
=\delta z+(1-\kappa_\star)\Omega_{mn}-f(\delta\z)
\end{equation}
where $\delta\z=\overline{\delta z}$. Here, we use the quasiperiodic
relation of $f(z)$ [which is inherited from that of $\zeta(z)$] and
some additional properties of the Weierstrass zeta function. Note also
that $\kappa_\star=\pi/(2\omega)^2$, $\kappa_\star=\pi/(2\sqrt3\omega^2)$, and
$\kappa_\star=\pi/(3\sqrt3\omega^2)$ for square, triangular, and honeycomb
lattices with side length of $\omega$. Assuming $\abs{\delta z}\ll
\omega$, equation (\ref{eq:ape}) can be solved approximately from
\begin{equation}
(1-\kappa_\star)\Omega_{mn}-\eta=f(\delta\z)-\delta z
\simeq\frac1{\delta\z}-\delta z+\mathscr R
\end{equation}
where the remainder term ($\mathscr R$) is the order of $(\delta\z)^j$
with $j=3$ (square), $j=5$ (triangle), or $j=2$ (hexagon).
Provided that $\abs{\mathscr L}\gg\omega^{-1}$ where $\mathscr L=
\abs{\mathscr L}\e^{\ij\phi_L}=(1-\kappa_\star)\Omega_{mn}-\eta$,
we have the image at $z=\Omega_{mn}+\delta z$ where
\begin{equation}
\label{eq:app}
\delta z\simeq\frac{\e^{\ij\phi_L}}{\abs{\mathscr L}}
\left(1-\frac1{\abs{\mathscr L}^2}\right)
+\mathscr O\biggl(\frac1{\abs{\mathscr L}^{\min(j+2,5)}}\biggr)
\end{equation}
and its signed magnification is found to be
\begin{equation}
\begin{split}
\mu=\mathscr J^{-1}=\frac1{1-\abs{f'(z)}^2}
=\frac1{1-\abs{f'(\delta z)}^2}
\\\simeq-\frac1{\abs{\mathscr L}^4}+\frac4{\abs{\mathscr L}^6}
+\mathscr O\biggl(\frac1{|\mathscr L|^{\min(j+5,8)}}\biggr).
\end{split}
\end{equation}
Since $\Omega_{mn}-\hat\eta=\mathscr L/(1-\kappa_\star)$ where
$\hat\eta=\eta/(1-\kappa_\star)$ with $\kappa_\star\ne1$, the condition that
$\abs{\mathscr L}>L_0$ is sufficiently large translates to the lens
positions ($\Omega_{mn}$) that lie outside the circular region of
radius $L_0/(1-\kappa_\star)$ centred at $\hat\eta$.\footnote{\scriptsize 
Note that,
as $\kappa_\star\rightarrow1$, the radius of the region within which the
approximation is not valid also increases.} Then, the images near all
of those lens positions are simply found by the approximation
(eq.~[\ref{eq:app}]). More importantly, the contribution to the total
magnification from all of those images is approximately
\begin{equation}
\begin{split}
\widehat{\sum\limits_{m,n}}\abs{\mu}\simeq
\widehat{\sum\limits_{m,n}}\frac1{\abs{\mathscr L}^4}
&\approx\int_{L_0/(1-\kappa_\star)}^\infty\!
\frac{2\pi\sigma R\,\df R}{\abs{\mathscr L}^4}
\\=&\int_{L_0/\abs{1-\kappa_\star}}^\infty\!
\frac{2\kappa_\star R\,\df R}{R^4(1-\kappa_\star)^4}
=\frac{\kappa_\star}{L_0^2(1-\kappa_\star)^2}
\end{split}
\end{equation}
where the summation is over the lens positions $(m,n)$ such that
$\abs{\Omega_{mn}-\hat\eta}>L_0/\abs{1-\kappa_\star}$ and
$\sigma=\kappa_\star/\pi$ is the number density of the lenses. Here the
polar coordinate centred at $\hat\eta$ is used for the integral and
so $R=\abs{\Omega_{mn}-\hat\eta}$. The result indicates that, unless
$\kappa_\star=1$, the contribution from infinitely many negative parity
images to the total magnification is finite\footnote{\scriptsize The sum
$\sum_{m,n}'\abs{\mathscr L}^{-r}$ where the summation is over the
lattice points except the points at which $\mathscr L=0$, is
`absolutely convergent' for $r>2$. For the rigorous proof, see the
mathematical references on elliptic functions listed in Appendix.}
(note that the integrand is convex so that the sum is bounded by the
integral) and with sufficiently large $L_0$, they are negligible.

In Figs.~\ref{figim1}-\ref{figimt}, we show the image locations
for some specific cases of lattice lenses. Here, all
positive parity images are located through the iteration scheme
outlined earlier. Most of negative parity images are found via
the Newton-Raphson method (eq.~[\ref{eq:sec2}]) with the starting
locations given by the approximate solutions in equation (\ref{eq:app}).
Additional locations of negative parity images are also found from
the iteration given by equation (\ref{eq:sec2}) with starting points
given by a grid of positions near $z=\eta/(1-\kappa_\star)$.

\begin{figure}
\includegraphics[width=\hsize]{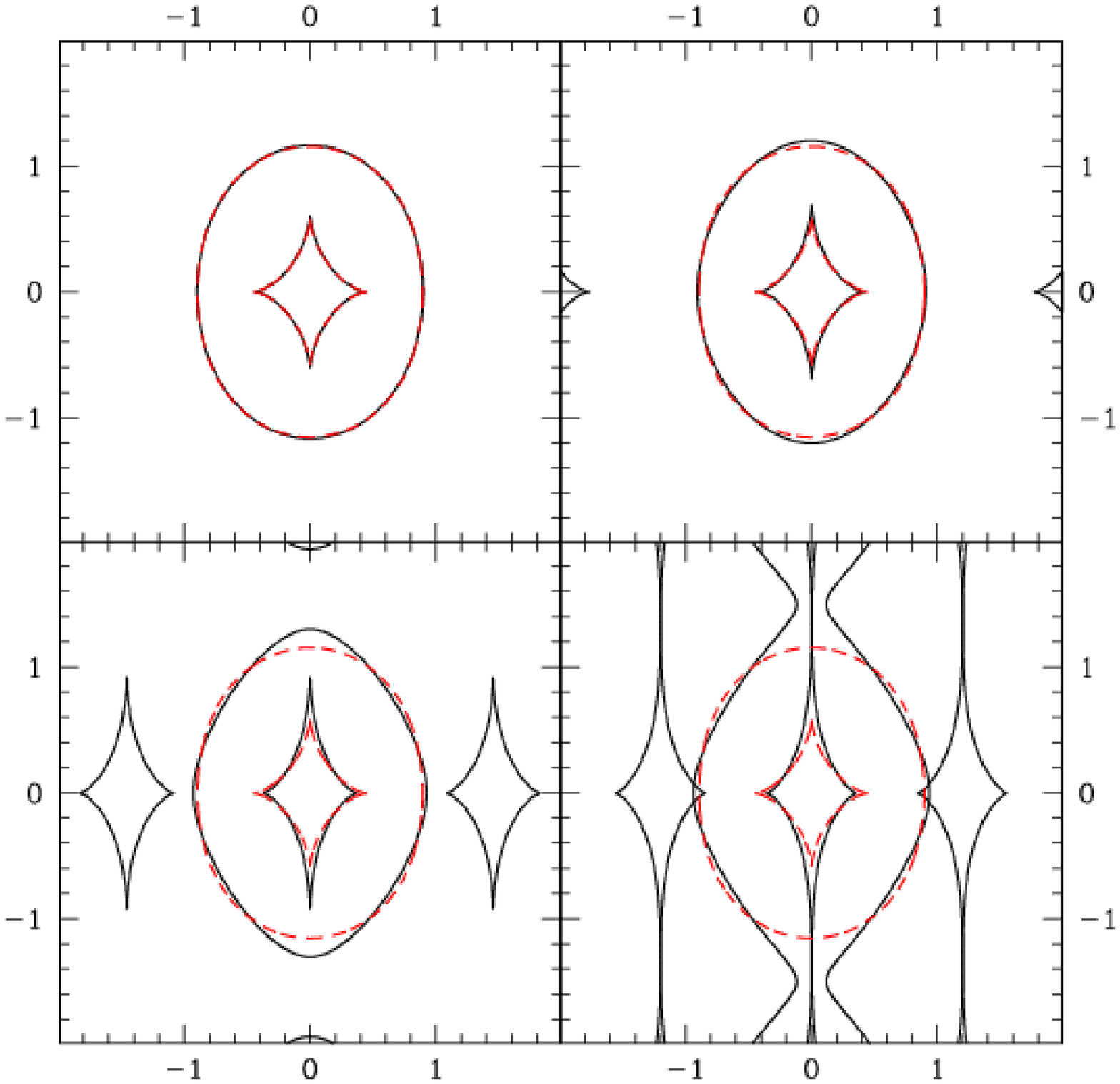}
\caption{\label{fig6} Critical curves and caustics for square
lattice lenses with external shear (solid lines). Also plotted in dashed
lines are the critical curves and caustics of the Chang-Refsdal lens
(a single point-mass lens) with the same external shear. All four
panels are for $\abs{\gamma_c}=0.25$, which is oriented parallel to the
side of the lattice cell, but $\kappa_\star$ varies; 0.1 (top left), 0.2
(top right), 0.3 (bottom left), and 0.35 (bottom right).}
\end{figure}

\section{External Shear}
\label{sec:sh}

Next, we consider the effect of the large-scale potential.
It is a usual practice that the external potential
$\psi_\mathrm{ext}$ in equation (\ref{eq:leq})
is approximated by a quadratic function and
consequently the corresponding deflection by a linear function (i.e.,
tensor) -- that is, keeping only the linear terms in its Taylor-series
expansion \citep{Yo81,NO84,SCN}. The resulting linear function is
symmetric, thanks to the mixed coordinate differential being
commutative. It is also customary that the linear function is
decomposed into a scalar dilation and a traceless tensor. The former,
usually referred to as a convergence, is directly related to the
surface mass density (which is not in the form of point masses)
whereas the latter, known as an external shear, physically models the tidal
effects from the external mass distribution.

In complex number notation, then the lens equation generalizes to
\begin{equation}
\label{eq:lcs}
\eta=(1-\kappa_c)z-\gamma_c\z-\overline{f(z)}\,;\qquad
\mathscr J=
(1-\kappa_c)^2-\Bigl\lvert\bigl[-f'(z)\bigr]-\bar\gamma_c\Bigr\rvert^2
\end{equation}
where $\kappa_c$, which must be positive real, is the convergence due
to the local surface mass density that is the source of the large
scale potential (i.e., `continuous mass density') and $\gamma_c$,
which can be any complex number, is the shear that models the tidal
effect. The effect of the convergence term is uniform focusing,
which basically results in a scaling difference between the image and
source planes. It is usually incorporated into the analysis by the
redefinition of the variables \citep{Pa86}. For example, with the
rescaled position variables for the image plane
$w=\abs{\epsilon}^{1/2}z$ and the source plane
$\xi=\sgn(\epsilon)\abs{\epsilon}^{-1/2}\eta$ where
$\epsilon=1-\kappa_c$, equation (\ref{eq:lcs}) for the square lattice
with side length of $\omega$ [i.e.,
$f(z)=\zeta(z;g_2,0)=\zeta(z\vert\omega,\ij\omega)$] reduces to
\begin{equation}
\begin{split}&
\xi=w-\varsigma\bar w
-\sgn(\epsilon)\ \overline{\zeta\bigl(w\vert\,
\abs{\epsilon}^{1/2}\omega,\ij\abs{\epsilon}^{1/2}\omega\bigr)}
\,;\\&
\frac{\mathscr J}{(1-\kappa_c)^2}=
1-\Bigl\lvert\wp\bigl(w\vert\,
\abs{\epsilon}^{1/2}\omega,\ij\abs{\epsilon}^{1/2}\omega\bigr)
-\sgn(\epsilon)\bar\varsigma\Bigr\rvert^2
\end{split}
\end{equation}
where $\varsigma=\gamma_c/\epsilon$ is the reduced shear. We note that
the deflection function is basically given by that for the square
lattice with side length of $\abs{\epsilon}^{1/2}\omega$ (i.e.,
corresponding to the reduced optical depth $\varkappa=\kappa_\star/\epsilon$).
In general, if $\epsilon>0$ (i.e., $0\le\kappa_c<1$), the qualitative
lensing behaviour is basically same as the case that $\kappa_c=0$
except the additional scaling difference between the image and the
source planes, which leads to the use of the `reduced' values of the
external shear and the optical depth \citep{Pa86,GSW,AE06}. For the
following discussion, we only consider the lens equation
(\ref{eq:lcs}) for the case $\kappa_c=0$.

\input f16.cap

The critical curves and the caustic networks of the lattice lens with
external shear can be found in the same way as the case without
external shear. However, the shear introduces two further degrees of
freedom in the parameter space to be explored, which makes the task
quite extensive and difficult to generalize. Nevertheless, we can draw
a rough generalization such that the patterns reduce to the case of
the Chang-Refsdal lens \citep{CR79,CR84,AE06} if
$\abs{\gamma_c}\gg\kappa_\star$ and to the simple lattices with no (or
negligible) shear if $\abs{\gamma_c}\ll\kappa_\star$.
Figs.~\ref{fig6}-\ref{fig8} show some examples of the critical curves and
the caustic networks for the square lattices with moderate values of
$\abs{\gamma_c}$ and $\kappa_\star$ to illustrate the effects of varying
$\kappa_\star$ at fixed $\gamma_c$ or varying $\gamma_c$ (both the magnitude
and the orientation) at fixed $\kappa_\star$. The situation becomes quite
complicated if $\abs{\gamma_c}\approx\kappa_\star\approx1$. As in the case of
lattices with zero shear, with $\kappa_\star\approx1$, there are overlaps among
neighboring caustics. Furthermore, with nonzero shear, neighboring
caustics can be connected with one another, and the resulting dense
networks of caustics are even more complicated to analyze
properly. More detailed analysis of these will be attempted in a
subsequent work.

Instead, here we speculate on some possible connections of lattice
lensing to more realistic lensing scenarios (particularly, that of the
random star field). One of the most significant differences of
lattice lensing from that by a random field is the fact that the
shear experienced at any given lens position due to the remaining
lenses cancels out to be null. Adding external shear only alters the
situation minimally, that is, the total shear at any lens position is
shifted to the value of external shear but every lens still
experiences the identical shear. By contrast, in the random star
field, the shear experienced by a given star is distributed according
to the same distribution as that of the shear experienced by the
random lines of sight. However, we note that the latter distribution
has been calculated \citep{NO84,Sc87,GSW} using the straightforward
application of the characteristic function in the probability theory
and the random-walk process. If lattice lensing were to model the
effect of the shear `variations' in the random field reasonably well,
each individual lens in the random field might be approximated as a
lattice lens with the `external shear' being equal to the actual total
shear experienced by the given lens. Then, the lensing behaviour of the
whole of the random star field could be understood as the sort of
`convolution' of the lattice lens (of the corresponding optical depth)
with the external shear distributed as the distribution of the random
shear in the random star field. If this indeed be the case, one could
study more realistic scenarios of the microlensing effects due to the
stars in the lensing galaxy in an alternative way to the traditional
inverse ray tracing approach.

\input f17.cap

At this point, however, this remains a speculative
possibility. As it has been noted widely, the effects of many lenses
combine in a highly nonlinear fashion and the role of the individual
lenses even in the simplest phenomenology is nontrivial
\citep[see e.g.,][]{GL05}. While
one could expect that the approach of decomposing the star field into
lattices outlined in the previous paragraph might fare better than the
simplistic approach of seeing the star field as the combination of
individual stars (since it takes care of some of the nonlinear effects),
no compelling argument can yet be made that this is the case.

\section{Conclusion}

Microlensing at high optical depth is not only of a theoretical interest
but also of an importance in understanding a number of lensed systems
\citep[e.g,][]{Ke06,Mo06}. However, at present time, the only way to
get any detailed quantitative grip on such problem is via Monte-Carlo
simulations of each individual point corresponding specific values of
the optical depth and the external shear \citep[e.g.,][]{GSW,SWL},
which is rather expensive in its requirement for resources. We argue that
a grid of point masses discussed in this paper, while artificial and
abstract so that it may appear to be entirely divorced from any physically
realistic problem, can provide us with some insight, albeit indirect, into
what is at work in microlensing at high optical depth, in particular
for the numbers and magnifications of microimages. Together with
with various analytical methods and probabilistic approaches to the
problem, this help us build pictures of phenomena complement to
more traditional routes with comparatively inexpensive means.

\section*{acknowledgments}
The author thanks P.~L.~Schechter for the initial stimulus of the project
and subsequent discussion and suggestions. The author is also grateful to
A.~Gould for careful reading of and detailed comments on the draft.
This work led the author's increased appreciation of remarkable geniuses
of C.~G.~Jacobi and K.~Weierstrass,
who have invented/discovered the theory of elliptic functions
over one and a half centuries ago. The author and
this work have been supported through the grant AST-0433809 from
the National Science Foundation (US).

\begin{appendix}

\section{Weierstrass Elliptic Functions}

Here, we summarize some basic concepts and properties
of elliptic functions and the Weierstrass functions
that are related
to the expositions given in the main text. For a general but concise
introduction to the subject, we refer to online resources such as
Wolfram Mathworld\footnote{\scriptsize \url{http://mathworld.wolfram.com/}}
or standard references on special functions such as \citet{AS}.
For more detailed and rigorous (and modern) treatments,
we advise our reader to look upon advanced texts on the complex analysis
\citep[e.g.,][]{Ah79,La99} or the analytic number theory
\citep[e.g.,][]{Ap97}, particularly those on the theory of the
elliptic curves and/or modular functions/forms
\citep[e.g.,][]{Si86,La87,Ko93}. In addition,
Large collections of formulae involving
the Weierstrass functions are also available online
(e.g., the Wolfram functions
site\footnote{\scriptsize \url{http://functions.wolfram.com/}}).

Elliptic functions are the class of complex-valued functions of a
complex variable that are meromorphic (i.e., analytic everywhere in
the complex plane except isolated poles) and biperiodic (or
`doubly-periodic'). The biperiodicity means that there exist two
complex numbers $\omega_1$ and $\omega_3$ such that
$\omega_1/\omega_3$ is not real and $f(z+2m\omega_1+2n\omega_3)=f(z)$
for all pairs of integers $m$ and $n$ and any complex number $z$.
Here, the two complex numbers $\omega_1$ and $\omega_3$ are referred to as
the half-periods of the elliptic function if they are a pair of two
`smallest' such numbers.\footnote{\scriptsize The reader should be warned that,
while we follow the convention of \citet{AS} in the paper, that is,
$\omega_i$ being half-periods, many modern mathematical references
adopt the different convention such that $\omega_i$ and the related
arguments represent the periods.} We note that the biperiodicity
further implies that, once the behaviour of the function is known in
the domain that is the parallelogram `cell' whose vertices are $0$,
$2\omega_1$, $2\omega_2\equiv2\omega_1+2\omega_3$ and $2\omega_3$ (or
equivalently $\omega_1-\omega_3$, $\omega_1+\omega_3$,
$-\omega_1+\omega_3$, $-\omega_1-\omega_3$), it is completely
specified everywhere in the complex plane. The cell is sometime
referred to as the fundamental period parallelogram (FPP) of the
elliptic function.
According to the general theory of elliptic
functions, any elliptic function can be expressed using a set of
standard functions, which are usually chosen to be one of two classes: Jacobi
functions ($\sn x$, $\cn x$, $\dn x$ and so on) or the Weierstrass
$\wp$-function and its derivative.

The Weierstrass $\wp$-function (or `the' Weierstrass elliptic
function) is the elliptic function that has one second-order pole in
its FPP. The standard definition locates a pole at $z=0$ whose principal
part is exactly $z^{-2}$ and constant part is nil (i.e.,
$\lim_{z\rightarrow0}[\wp(z)-z^{-2}]=0$). The formal definition is
given by
%
\[
\wp(z\vert\omega_1,\omega_3)\equiv\frac1{z^2}+
\sideset{}{'}\sum_{\substack{m,n=-\infty\\(m,n)\ne(0,0)}}^\infty
\left[\frac1{(z-\Omega_{mn})^2}-\frac1{\Omega_{mn}^2}\right]
\]
%
where $\Omega_{mn}=2m\omega_1+2n\omega_3$, and the summation is over
all integer grid points of $(m,n)$ except $(m,n)=(0,0)$. Here, the set
of all complex numbers in the form of $\Omega_{mn}$ is usually
referred to as a lattice. Note that the function $\wp(z)$ has
second-order poles at every point of the lattice. In addition,
$\wp(z)$ is an even function, i.e.,
$\wp(-z)=\wp(z)$. Note that, for a given lattice, the choice of the
pair of the half-periods is not unique -- that is to say, the
lattice forms a module over the ring of integers and the set of two
periods (i.e., $\{2\omega_1,2\omega_3\}$)
is the basis of this module.
For this reason, the two secondary arguments
of $\wp(z)$ are usually replaced by the elliptic invariants defined by
$g_2=60\sum_{m,n}'\Omega_{mn}^{-4}$ and
$g_3=140\sum_{m,n}'\Omega_{mn}^{-6}$ such that
$\wp(z;g_2,g_3)=\wp(z\vert\omega_1,\omega_3)$. Unlike a pair of the
half-periods, the elliptic invariants are in fact uniquely specified
by the given lattice, and vice versa.
 
The elliptic invariants are directly
related to the coefficients of the Laurent-series
expansion of $\wp$-function at origin via simple algebraic
relations, that is, the coefficients are given by the polynomials
of $g_2$ and $g_3$. Particularly, we have
$\wp(z;g_2,g_3)\simeq z^{-2}+(g_2/20)z^2
+(g_3/28)z^4+\mathscr O(z^6)$.
This series expansion further indicates that
$[\wp'(z;g_2,g_3)]^2-4[\wp(z;g_2,g_3)]^3-g_2\wp(z;g_2,g_3)
=g_3+\mathscr O(z^2)$. This combined with the results from the
general theory of elliptic functions implies that
$y(z)=\wp(z;g_2,g_3)$ is a particular solution of
the differential equation given by
$(y')^2=4y^3-g_2y-g_3$.
Since this is a first-order differential equation that does not
involve the independent variable, its general solution is given by
$y(z)=\wp(z+C;g_2,g_3)$ where $C$ is an integration constant (which is an
arbitrary complex constant). Note that,
if the polynomial $4y^3-g_2y-g_3$ can be factored with
a perfect square factor (so that $g_2^3=27g_3^2$), the
solution of the differential equation then can be expressed in terms
of `elementary' functions such as trigonometric or hyperbolic
functions and so does the Weierstrass function.
In particular, if
$g_2=12e^2>0$ and $g_3=-8e^3$ are both nonzero real [i.e.,
$4y^3-g_2y-g_3=4(y-e)^2(y+2e)$], then
\[
\wp(z;12e^2,-8e^3)=
\begin{cases}
3e\coth^2[(3e)^{1/2}z]-2e
&\text{if $e>0$}\\
(-3e)\cot^2[(-3e)^{1/2}z]+(-2e)
&\text{if $e<0$}
\end{cases}.
\]
However, this case
actually corresponds (at least, formally) to $\omega_1=\infty$ or
$\omega_3=\infty$, and the resulting functions are not elliptic but
simply (singly-)periodic [unless $g_2=g_3=0$, for which
$\wp(z;0,0)=z^{-2}$ and $\omega_1=\omega_3=\infty$ so that it is
in fact aperiodic].
  
The derivative of $\wp$-function
\begin{displaymath}
\wp'(z;g_2,g_3)=
\wp'(z\vert\omega_1,\omega_3)
\equiv\frac\df{\df z}\wp(z\vert\omega_1,\omega_3)
=-\sum_{m=-\infty}^\infty\sum_{n=-\infty}^\infty
\frac2{(z-\Omega_{mn})^3}
\end{displaymath}
is also an elliptic function with the same half-periods as
$\wp(z\vert\omega_1,\omega_3)$ but it has third-order poles at the
lattice points and is an odd function. By contrast,
the antiderivative
(or the indefinite integral) of $\wp$-function cannot be an elliptic
function. According to the general theory of elliptic functions, the
sum of complex residues in FPP of any elliptic function must vanish.
However, if the antiderivative of $\wp$-function were an elliptic
function, it would have a single simple pole in its FPP and thus the
sum of its residues could not vanish. In fact,
the Weierstrass
$\zeta$-function\footnote{\scriptsize The Weierstrass $\zeta$-function
should not be confused with several other
mathematical functions conventionally denoted by the same symbols,
the most important among which is the Riemann $\zeta$-function and
its generalization, the Hurwitz $\zeta$-function.},
which is defined to be a particular anntiderivative of
$\wp$-function;
\begin{displaymath}
\begin{split}
\zeta(z;g_2,g_3)=
\zeta(z\vert\omega_1,\omega_3)
=\frac1z+
\int_0^z\biggl[\frac1{w^2}-\wp(w\vert\omega_1,\omega_3)\biggr]\df w
\\=\frac1z+
\sideset{}{'}\sum_{m,n}
\left(\frac1{z-\Omega_{mn}}+\frac1{\Omega_{mn}}+\frac z{\Omega_{mn}^2}\right)
\end{split}
\end{displaymath}
is a quasiperiodic function such that
$\zeta(z+2m\omega_1+2n\omega_3)=\zeta(z)+2m\zeta(\omega_1)+2n\zeta(\omega_3)$
for all pairs of integers $m$ and $n$ and any complex number $z$ with
$\omega_1$ and $\omega_3$ being the half-periods of the corresponding
$\wp$-function (secondary arguments suppressed for brevity). We also
note that the Weierstrass $\zeta$-function is an odd function.

In addition, the theory of the Weierstrass functions also introduces
another auxiliary function referred to as the Weierstrass
$\sigma$-function;
\[
\sigma(z;g_2,g_3)=\sigma(z\vert\omega_1,\omega_3)
\equiv z
\sideset{}{'}\prod_{\substack{m,n=-\infty\\(m,n)\ne(0,0)}}^\infty
\left[\left(1-\frac z{\Omega_{mn}}\right)
\exp\left(\frac z{\Omega_{mn}}+\frac{z^2}{2\Omega_{mn}^2}\right)\right]
\]
where the product is again over all integer pairs $(m,n)$ except
$(m,n)=(0,0)$. From this definition, it is straightforward to show
that $\zeta(z)$ is the log-derivative of $\sigma(z)$, that is,
\[
\frac{\sigma'(z)}{\sigma(z)}=
\frac\df{\df z}\ln\sigma(z;g_2,g_3)=\zeta(z;g_2,g_3).
\]
We note that $\sigma(z)$ is also a quasiperiodic function although the
relation involves the successive ratios between the functional values
at the points separated by the `period', rather than the differences as in
$\zeta(z)$;
\[
\frac{\sigma(z+2m\omega_1+2n\omega_3)}{\sigma(z)}
=(-1)^p\exp\left\{2[m\zeta(\omega)+n\zeta(\omega)]
(z+m\omega_1+n\omega_3)\right\}
\]
where $p=m+n+mn$ and $m$ and $n$ are any pair of integers.
The function $\sigma(z)$ is in
fact an entire function, that is, it is analytic everywhere in the
complex plane without any pole or singularity (except the essential
singularity at infinity). In particular, it has zeros at the
origin and the lattice points, that is,
$\sigma(0)=\sigma(\Omega_{mn})=0$, which is easily verifiable from its
definition.

We also note that
the set of Weierstrass functions satisfy the homogeneity
relations, which can be easily derived from their respective
definitions in terms
of the infinite sum or product over the lattice points.
In particular,
\begin{align*}
&\sigma(az\vert a\omega_1,a\omega_3)=
a\sigma(z\vert\omega_1,\omega_3)\,;
&&\sigma(az;a^{-4}g_2,a^{-6}g_3)=a\sigma(z;g_2,g_3)\,,
\\
&\zeta(az\vert a\omega_1,a\omega_3)=
a^{-1}\zeta(z\vert\omega_1,\omega_3)\,;
&&\zeta(az;a^{-4}g_2,a^{-6}g_3)=a^{-1}\zeta(z;g_2,g_3)\,,
\\
&\wp(az\vert a\omega_1,a\omega_3)=
a^{-2}\wp(z\vert\omega_1,\omega_3)\,;
&&\wp(az;a^{-4}g_2,a^{-6}g_3)=a^{-2}\wp(z;g_2,g_3)\,,
\\
&\wp'(az\vert a\omega_1,a\omega_3)=
a^{-3}\wp'(z\vert\omega_1,\omega_3)\,;
&&\wp'(az;a^{-4}g_2,a^{-6}g_3)=a^{-3}\wp'(z;g_2,g_3)\,,
\end{align*}
\[
g_2(a\omega_1,a\omega_3)=a^{-4}g_2(\omega_1,\omega_3)\,,\qquad
g_3(a\omega_1,a\omega_3)=a^{-6}g_3(\omega_1,\omega_3)\,,
\]
where $a\ne0$ is a complex constant.

\begin{table*}
\begin{minipage}{122mm}
\caption{Laurent-Series Coefficients of $\wp(z)$ at $z=0$.}
\label{tab:co}
\begin{tabular}{ccc}
\hline
$k$&$a_{k+1}/a_k$&$b_{k+1}/b_k$\\
\hline
1&$1/3$&$1/13$\\
2&$2/13$&$1/19$\\
3&$5/(2\cdot17)=5/34$&$3/(5\cdot13)=3/65$\\
4&$2/(3\cdot5)=2/15$&$2^2/(3\cdot31)=4/93$\\
5&$5/(3\cdot13)=5/39$&
($5\cdot7\cdot43)/(2^2\cdot13\cdot19\cdot37)=1505/36556$\\
6&$(2\cdot3^2)/(5\cdot29)=18/145$&
$(2\cdot3\cdot431)/(5\cdot7\cdot43^2)=2586/64715$\\
\hline
\medskip
For both cases, $a_1=b_1=1$.
\end{tabular}
\end{minipage}
\end{table*}

\subsection{lemniscatic case}

If FPP is a square (i.e., $\omega_3=\ij\omega_1$), then
$\Omega_{mn}=2m\omega_1+2n\omega_3=2(m+n\ij)\omega_1$ and, from the
symmetry, we have $\sum_{mn}'\Omega_{mn}^{-6}
=(2\omega_1)^{-6}\sum_{mn}'(m+n\ij)^{-6}=0$, and so that $g_3=0$. Then,
after $g_2$ is reduced to the unity using the homogeneity relation,
the behaviour of the corresponding Weierstrass elliptic function can be
studied through that of $\wp(z;1,0)$, which is called the lemniscatic
case.
The corresponding half-periods of
$\wp(z;1,0)=\wp(z\vert\omega_0,\ij\omega_0)$ are given by
$\omega_0
=\Gamma^2_{1/4}/(4\sqrt{\pi})\approx1.854$
and $\ij\omega_0$. (The constant $\sqrt2\omega_0
\approx2.622$ is
sometimes known as the lemniscate constant.)
Some basic results known for the lemniscatic case are
\begin{align*}
&\wp(\omega_0;1,0)=\frac12\,;
&&\zeta(\omega_0;1,0)=\frac\pi{4\omega_0}\,,
\\&\wp(z_0;1,0)=0\,;
&&\zeta(z_0;1,0)=\frac\pi{4\omega_0}(1-\ij)\,,
\\&\wp(\ij\omega_0;1,0)=-\frac12\,;
&&\zeta(\ij\omega_0;1,0)=-\frac{\pi\ij}{4\omega_0}.
\end{align*}
Here, $z_0=(1+\ij)\omega_0=\sqrt2\e^{\pi\ij/4}\omega_0$ is in fact the
only zero of $\wp(z;1,0)$ within its FPP (which is actually a
degenerate second-order zero).
Note that these zeros are located at
the centres of the square cells defined by the poles. 
We further note that the behaviours of these functions within FPP exhibit
high degrees of symmetry and their study can be reduced to the
one-eighth of FPP, that is, the isosceles-right-triangular region defined by
$z=0$ (the pole), $\omega_0$ and $z_0$ (the zero).

Although the standard reference on the numerical calculation,
\citet{NR}, does not list the routine for the Weierstrass functions,
their numerical evaluation is available in commercial softwares such as
Mathematica$^\text\textregistered$ or Maple$^\text\texttrademark$
or subroutine libraries such as
IMSL$^\text\texttrademark$\footnote{\scriptsize \url{http://www.vni.com/products/imsl/}}.
However, for most purposes,
they can be evaluated in a pretty straightforward manner without
using any black-box routine \citep[see e.g.,][]{AS}.
While there exist more involved algorithms that are valid for an
arbitrary pair of complex numbers $(g_2,g_3)$, the lemniscatic case of
the Weierstrass function can be easily evaluated through its
Laurent-series expansion at $z=0$ or Taylor-series expansion at
$z=\omega_0$ or $z=z_0$ (after the reduction of the domain using
biperiodicity and additional symmetry properties). The relevant series
expansions up to the first several terms are available in
\citet{AS}. In particular, the Laurent series at $z=0$ is given in the
form of $\wp(z;1,0)=z^{-2}[1+\sum_{k=1}^\infty(a_k/20^k)z^{4k}]$ (the
first few coefficients $a_k$'s of which are given in Table
\ref{tab:co}) so that it converges relatively quickly for
$\abs{z}<1$ -- the radius of the convergence of the series is in fact
$2\omega_0$. Note that the coefficients for the Laurent series at
$z=0$ can in fact be derived using a total recurrence relation.

Since the lemniscatic case of $\wp$-function
satisfies the differential equation $(y')^2=4y^3-y$,
its inverse function is expressible through elliptic integrals.
However, the relevant integral in fact more easily reduces to
an incomplete beta function
\[
z=\wp^{-1}(\wp;1,0)
=-\int_\infty^\wp\!\frac{\df y}{\sqrt{4y^3-y}}
=\frac1{2^{3/2}}\ B_{1/(4\wp^2)}\biggl(\frac14,\frac12\biggr)
\]
or a hypergeometric function,
\[
z=\wp^{-1}(\wp;1,0)
=\frac1{\wp^{1/2}}\
{}_2F_1\biggl(\frac12,\frac14;\frac54;\frac1{4\wp^2}\biggr)
\]
%
where $B_x(a,b)$ is the incomplete Beta function and
${}_2F_1(a,b;c;x)$ is the Gaussian $(2,1)$-hypergeometric function.
[c.f., $B_x(a,b)=(x^a/a){}_2F_1(1-b,a;a+1;x)$.]
Here, the last
expression may be used for a convergent hypergeometric power-series
expansion for the inverse $\wp$-function for $\abs{\wp}\gg2^{-1}$.
We note that, because we are interested in the case when $\wp$ takes
an arbitrary complex value, there is no particular advantage using
the Legendre/Jacobi elliptic integrals or the incomplete Beta function,
numerical or otherwise, compared to the hypergeometric series expression 
given above.
Alternative convergent
hypergeometric series for
$\abs{\wp}\ll2^{-1}$ also exists;
%
\begin{displaymath}
z=
\wp^{-1}(\wp;1,0)
=z_0+2\ij\wp^{1/2}\ {}_2F_1\biggl(\frac12,\frac14;\frac54;4\wp^2\biggr)
\end{displaymath}
%
where $z_0=\sqrt2\e^{\pi\ij/4}\omega_0$ is again the zero of $\wp(z;1,0)$.
The inverse function can be effectively evaluated almost everywhere
using these
hypergeometric series
together with some further
refinement methods such as the (complex) Newton-Raphson algorithm if
necessary.

\subsection{equiharmonic case}

We note that FPP can be considered as a tile, the set of which covers
the whole (complex) plane and the locations of the poles are vertices of
individual tiles. Then, the preceding lemniscatic case corresponds to the
tile given by a square (i.e, a regular tetra-gon), which is one of the
three possible regular tessellations of the plane.
The elliptic function with FPP corresponding to the tile pattern of
the remaining two regular tessellations can also be studied
similarly. The half-periods for the more basic of the two, that is,
the tiles given by equilateral triangles, are related to each other
via $\omega_3=\e^{\pi\ij/3}\omega_1$ and then from the symmetry of the
system, it follows that $g_2=60{\sum_{mn}}'\Omega_{mn}^{-4}=0$ (note
that, while the choice of the two half-periods is not unique, the
elliptic invariants are fixed for the given arrangement of poles). As
before, the remaining invariant $g_3$ can be reduced to the unity
using the homogeneity relation, and the resulting Weierstrass elliptic
function, $\wp(z;0,1)$ is referred to as the equiharmonic case.
It is known that the real half-period
of $\wp(z;0,1)=\wp(z\vert\omega_2,\e^{\pi\ij/3}\omega_2)$ is
$\omega_2=\Gamma^3_{1/3}/(4\pi)\approx1.530$.
Additionally, it can be derived that
\begin{align*}
&\wp(\e^{-\pi\ij/3}\omega_2;0,1)=2^{-2/3}\e^{2\pi\ij/3}\,;
&&\zeta(\e^{-\pi\ij/3}\omega_2;0,1)=\frac{\pi\e^{\pi\ij/3}}{2\sqrt3\omega_2}\,,
\\&\wp(\omega_2;0,1)=2^{-2/3}\,;
&&\zeta(\omega_2;0,1)=\frac\pi{2\sqrt3\omega_2}\,,
\\&\wp(\e^{\pi\ij/3}\omega_2;0,1)=2^{-2/3}\e^{-2\pi\ij/3}\,;
&&\zeta(\e^{\pi\ij/3}\omega_2;0,1)=\frac{\pi\e^{-\pi\ij/3}}{2\sqrt3\omega_2}\,,
\\&\wp(\sqrt3\ij\omega_2;0,1)=2^{-2/3}\,;
&&\zeta(\sqrt3\ij\omega_2;0,1)=-\frac{\pi\ij}{2\omega_2}.
\end{align*}
The zeros of $\wp(z;0,1)$ are found at the centres of each equilateral
triangular cells so that there are actually two zeros within its
FPP, and unlike the lemniscatic case, each zero is simple (i.e., the
first-order zero). The particular zero within the cell defined by
$z=0$, $z=2\omega_2$, and $z=2\e^{\pi\ij/3}\omega_2$ is located at
$z_0=(2/\sqrt3)\e^{\pi\ij/6}\omega_2$ with
$\zeta(z_0;0,1)=\pi\e^{-\pi\ij/6}/(3\omega_2)$.
%

Like the lemniscatic case, the equiharmonic case of the
Weierstrass function can also be readily evaluated via the
power-series expansions at $z=0$, $z=\omega_2$, or $z=z_0$. Their
coefficients are again found in the references such
as \citet{AS}. In fact, the Laurent-series expansion at $z=0$ of the
equiharmonic case (whose radius of convergence is $2\omega_2$)
is in the form of
$\wp(z;0,1)=z^{-2}[1+\sum_{k=1}^\infty(b_k/28^k)z^{6k}]$, which
converges even faster than that of the lemniscatic case for $\abs{z}<1$.
The first few coefficient $b_k$'s are again given in Table \ref{tab:co}.

Analogous to
the lemniscatic case, the inverse of $\wp$-function can be obtained
through the integration of the differential equation $(y')^2=4y^3-1$.
The resulting expression can again be written down
in terms of elliptic integrals (which is rather complicated),
incomplete beta functions, or
hypergeometric functions,
\begin{displaymath}
\begin{split}
z=\wp^{-1}(\wp;0,1)
=-\int_\infty^\wp\!\frac{\df y}{\sqrt{4y^3-1}}
=\frac1{2^{2/3}\cdot3}\ B_{1/(4\wp^3)}\biggl(\frac16,\frac12\biggr)
\\
=\frac1{\wp^{1/2}}\
{}_2F_1\biggl(\frac12,\frac16;\frac76;\frac1{4\wp^3}\biggr),
\end{split}
\end{displaymath}
%
Similarly, the last expression can be used
for the convergent hypergeometric power-series
expansion of the inverse $\wp$-function for $\abs{\wp}\gg2^{-2/3}$.
For the case when $\abs{\wp}\ll2^{-2/3}$, an alternative
expression involving a hypergeometric function
can be derived as before;
\begin{displaymath}
z=\wp^{-1}(\wp;0,1)
=z_0-\ij\wp\
{}_2F_1\biggl(\frac12,\frac13;\frac43;4\wp^3\biggr).
\end{displaymath}
%
Here, $z_0=(2/\sqrt3)\e^{\pi\ij/6}\omega_2$ is again the zero
of $\wp(z;0,1)$. The sign in front of the hypergeometric function is
actually dependent upon the choice of $z_0$. For example, with an
alternative zero $z_0=-(2/\sqrt3)\e^{\pi\ij/6}\omega_2$, it should be
positive. Again, these hypergeometric series supplemented
by the Newton-Raphson algorithm provide an efficient way to evaluate
the inverse $\wp$-function for the equiharmonic case.

\subsubsection{hexagonal grid}
\label{app:hex}

Let us consider an elliptic function,
\begin{equation}
\label{eq:hex}
\mathfrak h(z;g_3)=\wp(z;0,g_3)-\wp\Bigl(z;0,-\frac{g_3}{27}\Bigr).
\end{equation}
Note that every pole of $\wp(z;0,-g_3/27)$ coincides with that of
$\wp(z;0,g_3)$ and they cancel each other leaving a fourth-order zero
at the location. On the other hand, the remaining poles of $\wp(z;0,g_3)$
are located at the vertices of regular hexagonal tiles.
Hence, equation (\ref{eq:hex}) corresponds to the remaining case of
the elliptic function with their (second-order) poles forming vertex
points of the regular tessellation of the space.

While it is a straightforward calculation,
equation~(\ref{eq:hex}) is unsuitable for the accurate numerical
evaluation near $z=0$ (and also near the other zeros of the functions)
because of the canceling of two divergent terms leading to the zero.
Furthermore, the inversion of equation~(\ref{eq:hex}) is also rather
nontrivial.
This difficulty can be overcome as follows.
We first note that
%
%
\[\begin{split}
\bigl[\mathfrak q'(z)\bigr]^2
&=\bigl[\mathfrak p'(z)\bigr]^2\,
\biggl(1-\frac{2g_3}{27\bigl[\mathfrak p(z)\bigr]^3}\biggr)^2
\\&=\biggl(4\bigl[\mathfrak p(z)\bigr]^3+\frac{g_3}{27}\biggr)\,
\biggl(1-\frac{2g_3}{27\bigl[\mathfrak p(z)\bigr]^3}\biggr)^2
=4\bigl[\mathfrak q(z)\bigr]^3-g_3
\end{split}\]
where $\mathfrak q(z)=\mathfrak p(z)+(g_3/27)[\mathfrak p(z)]^{-2}$ and
$\mathfrak p(z)=\wp(z;0,-g_3/27)$.
In other words, $\mathfrak q(z)$ is the solution
of the differential equation $(y')^2=4y^3-g_3$, whose general solution
is in the form of $y=\wp(z+C;0,g_3)$ where $C$ is the (complex) integration
constant. Since the principal part of the Laurent-series expansion
of $\mathfrak q(z)$ at $z=0$ is $z^{-2}$, it is in fact the case that
$\mathfrak q(z)=\wp(z;0,g_3)$, and
consequently we have
%
\begin{equation}
\label{eq:hfun}
\mathfrak h(z;g_3)
=\frac{g_3}{27}\biggl[\wp\Bigl(z;0,-\frac{g_3}{27}\Bigr)\biggr]^{-2}
=\frac1{u^4}\biggl(\frac{g_3}{27}\biggr)^{1/3}\Biggl[\wp\biggl(
u\Bigl(\frac{g_3}{27}\Bigr)^{1/6}z;0,1\biggr)\Biggr]^{-2}
\end{equation}
where $u^6=-1$. Subsequently, we find that the inverse function
of equation (\ref{eq:hex}) or equivalently that of equation (\ref{eq:hfun})
is given by
\[
z=\frac{\mathfrak h^{1/4}}{(g_3/27)^{1/4}}\
{}_2F_1\biggl(\frac12,\frac16;\frac76;
-\frac{\mathfrak h^{3/2}}{4(g_3/27)^{1/2}}\biggr)
\]
or
\[
z=\frac{\Gamma^3_{1/3}}{2\pi g_3^{1/6}}-\frac1{\mathfrak h^{1/2}}
{}_2F_1\biggl(\frac12,\frac13;\frac43;
-\frac{4(g_3/27)^{1/2}}{\mathfrak h^{3/2}}\biggr)
\]
where some of signs are chosen so that the principal value
for positive real values of $\mathfrak h$ returns with being positive
real (in particular, $\mathfrak h\ge0\ \rightarrow\ z\in[0,2\omega)$).

On the other hand, if one wants to find the Taylor-series expansion of
$\mathfrak h(z;g_3)$ at $z=0$, it may be easier to use 
equation (\ref{eq:hex}) at $z=0$ directly;
\[\begin{split}
\mathfrak h(z;g_3)
&=\sum_{k=1}^\infty\left[1-\left(-\frac1{27}\right)^k\right]
\left(\frac{g_3}{28}\right)^kb_kz^{6k-2}
\\&=\frac{g_3}{27}z^4\left\{1+\sum_{k=1}^\infty
\left[\sum_{p=0}^k(-1)^{k-p}27^p\right]
b_{k+1}\left(\frac{g_3}{27\cdot28}z^6\right)^k\right\}
\end{split}\]
where $b_k$ is the Laurent-series coefficient of the equiharmonic
case of $\wp$-function (see Table \ref{tab:co}).
In particular, $b_1=1$ which is
actually used for the second equality of the preceding equation.

\end{appendix}

\bsp
\label{finish}
\end{document}